\documentclass[showpacs,amsmath,amssymb,graphicx,12pt]{revtex4}
\usepackage{graphicx}
\usepackage{epsfig}
\usepackage{multirow}
\usepackage{subfigure}
\usepackage{dcolumn}
\usepackage{bm}
\begin{document}

\title{$J/\psi \rightarrow p
\bar{p}\phi$ decay in the isobar resonance model}
\author{Jian-Ping Dai$^{1,4}$} \email{
daijianping@mail.ihep.ac.cn}
\author{Peng-Nian Shen$^{1,2,5,6}$} \email{
shenpn@mail.ihep.ac.cn}
\author{Ju-Jun Xie$^{2,3}$} \email{
xiejujun@mail.ihep.ac.cn}
\author{Bing-Song Zou$^{1,2,4,5}$} \email{
zoubs@ihep.ac.cn}
\affiliation{1 Institute of High Energy Physics,
CAS,
Beijing 100049, China\\
2 Theoretical Physics Center for Science Facilities, CAS,
Beijing 100049, China\\
3 Department of Physics, Zhengzhou University, Zhengzhou,
Henan 450001, China\\
4 Graduate University of Chinese Academy of Sciences,
Beijing 100049, China\\
5 Center of Theoretical Nuclear Physics, National
Laboratory of Heavy\\
Ion Accelerator, Lanzhou 730000, China\\
6 College of Physics and Technology, Guangxi Normal University,
Guilin  541004, China}

\begin{abstract}
Based on the effective Lagrangian approach, the $ J/\psi \to p
\bar{p} \phi$ decay is studied in an isobar resonance model with the
assumption that the $\phi$-meson is produced from intermediate
nucleon resonances. The contributions from the $N^*_{1/2^-}(1535)$,
$N^*_{3/2^+}(1900)$, $N^*_{1/2^-}(2090)$ and $N^*_{1/2^+}(2100)$
states are considered. In terms of the coupling constants
$g^{2}_{\phi N N^{*}}$ and $g^{2}_{\phi N N^{*}}$ extracted from the
data of the partial decay widths of the $N^*$s to the $N\pi$
channel, the reaction cross section of the $\pi^{-}p\rightarrow
n\phi$ process and the partial decay widths of the
$J/\psi\rightarrow p\bar{p}\eta$ and $J/\psi\rightarrow
p\bar{n}\pi^{-}$ processes, respectively, the invariant mass
spectrum and the Dalitz plot for $ J/\psi \to p \bar{p} \phi$ are
predicted. It is shown that there are two types of results. In the
type I case, a large peak structure around 2.09GeV implies that a
considerable mount of $N\phi$ or $qqqs\bar s$ component may exist in
the narrow-width $N^*_{1/2^-}(2090)$ state, but for the wide-width
$N^*_{1/2^+}(2100)$ state, it has little $qqqs\bar s$ component. In
the type II case, a small peak around 2.11GeV may only indicate the
existence of a certain mount of $p\phi$ or $qqqs\bar s$ component in
the narrow-width $N^*_{1/2^+}(2100)$ state, but no information for
the wide-width $N^*_{1/2^-}(2090)$ state. Further BESIII data with
high statistics would help us to distinguish the strange structures
of these $N^*$s.
\end{abstract}

\maketitle{}
\section{Introduction}
%
In past decades, many excited states of nucleon were observed and
their properties, such as the mass, width, decay modes, decay
branching fractions and etc., were more or less accurately measured.
Most of these states and their properties can be well-explained by
quark models, but some of them cannot be fitted into the nucleon
spectrum predicted by the three-valence-quark model. To explain the
discrepancy, except that the data is lack of higher accuracy and
statistics due to the limited experimental technique and method, one
speculated that these states may contain some constituents other
than three $u$ and $d$ valance quarks, especially the $s$ and
$\bar{s}$ quarks, and suggested to check this conjecture through
experiments. Later, in the high energy physics and nuclear physics
experiments, through the data analysis, one found that some excited
states of nucleon ($N^*$) couple strongly with strange particles.
For instance, in either $J/\psi\rightarrow\bar{p}K^{+}\Lambda$
decay, or $p p\rightarrow p\Lambda K^{+}$ reaction near the
kaon-production threshold\cite{b.ch.liu1,b.ch.liu2}, or $\gamma
p\rightarrow K^{+}\Lambda$ kaon-photoproduction
process\cite{m.q.tran, J.W.C.MCNABB,g.penner,b.julia-diaz},
$N^{*}(1535)$ has a significant strength of coupling to the
$K\Lambda$ channel. This indicates that the $N^{*}(1535)$ state may
contain a considerable mount of $s\bar{s}$ component, which is
consistent with a very large branching fraction of $45\sim 60$\% for
the $N^{*}(1535)\to N\eta$ decay.

On the other hand, the $\phi$-meson is mainly composed of
$s\bar{s}$. According to the Okubo-Zweig-Iizuka (OZI)
rule~\cite{s.okubo}, the production rate of $\phi$-meson in the
nuclear process would be suppressed if the initial interacting
particles do not contain a constituent with $s$ and $\bar{s}$
quarks. On the contrary, if a $N^*$ contains strange constituents,
its coupling with a channel involving a $\phi$-meson might be
relatively strong. In fact, it is found that the $pp\rightarrow
pp\phi$ and $\pi^- p\rightarrow n\phi$ reaction data can be
well-explained as long as the coupling constant of $\phi N
N^{*}(1535)$ is sufficiently large, which implies that such a
significant coupling is closely related to a fact that a
considerable mount of $s\bar{s}$ component is involved in the wave
function of $N^{*}(1535)$~\cite{j.j.xie}. Therefore, in the charm
physics experiment at BESIII, say the measurements of $J/\psi$
hadronic decays, the $N\phi$ decay channel of $J/\psi$ would also be
a good place to check whether some $N^{*}s$, as the intermediate
states in the decay process, have strange components, although the
branching fractions of such decays are not large.

Similar to $N^{*}(1535)$, some nucleon resonances which have not yet
been well-established and cannot be fitted into the nucleon spectrum
from theoretical models have remarkable branching fractions in some
decay channels involving strange particles, say $N\eta$, $\Lambda K$
and etc. For instance, the branching fractions for $N^*(2090)\to
N\eta$, $N^*(2100)\to N\eta$ and $N^*(1900)\to N\eta$ are about
41\%, 61\% and 14\%, respectively~\cite{pdg2008}. This implies that
these states might have sizable strange constituents, and the effect
of such ingredients should show up in the $J/\psi\rightarrow p
\bar{p} \phi$ decay.

In fact, the branching fraction of $J/\psi\rightarrow p \bar{p}\phi$
was measured by the DM2 Collaboration in 1988~\cite{J.E. Augusttin}.
However, due to the insufficient statistics, no resonance
information was extracted. Recently, the luminosity of BEPCII has
reached over $3\times10^{32}cm^{-2}s^{-1}$ around $J/\psi$ peak, a
huge amount of $J/\psi$ events, say $3\times10^{9}$, will be
collected at BESIII in one year. The new data set would offer an
opportunity to study the possible strange ingredient or even
pentaquark in the nucleon resonance.

Based on the mostly accepted assertion that the $J/\psi\rightarrow p
\bar{p}\phi$ decay is dominated by a process with intermediate
nucleon resonances, so-called resonance model, we study the
possibility of strange ingredients in the mentioned resonances
through this decay in an effective Lagrangian approach. It is our
hope that the information of the strange structure in nucleon
resonances, especially those which are not well-established, can be
deduced, and a reference for coming BESIII data analysis can be
provided.

The paper is organized in the following way. In Section II, the
theoretical model and formalism are briefly introduced. The results
are presented and discussed in Section III. And in Section IV, a
concluding remark is given.

\section{Model and Formalism}
\begin{figure}[htbp]
\begin{center}
\includegraphics[scale=0.8]{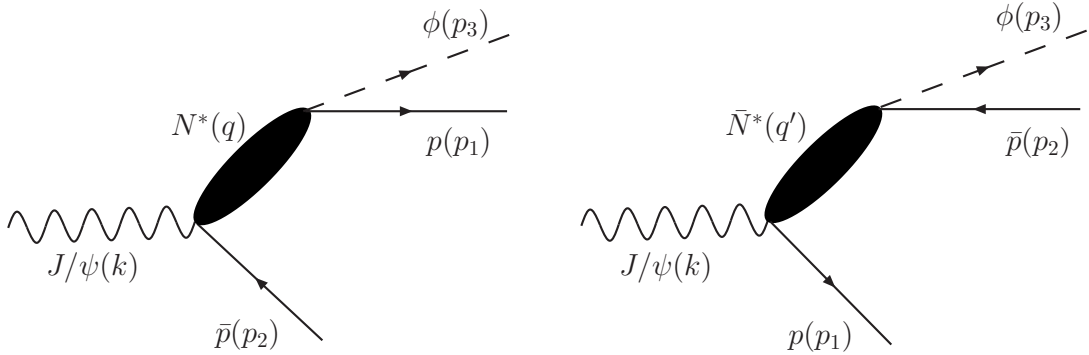} \caption{Feynman diagrams for the
$J/\psi\rightarrow p\bar{p}\phi$ decay in the resonance
model.} \label{ppbarphi}
\end{center}
\end{figure}

In the resonance model, the $J/\psi\rightarrow p \bar{p}\phi$ decay
undergoes a two-step process, namely $J/\psi$ firstly decays into an
intermediate $\bar{p}N^*$ ($p\bar{N^*}$) state, and then $N^*$
($\bar{N^*}$) successively decays to $\phi$ and $p$ ($\bar{p}$).
Corresponding Feynman diagrams are drawn in Fig.\ref{ppbarphi},
where $k$, $p_{1}$, $p_{2}$, $p_{3}$ and $q$($q'$) are the
four-momenta of $J/\psi$, $p$, $\bar{p}$, $\phi$ and
$N^{*}$($\bar{N^{*}}$), respectively.
The embedded intermediate $N^*$ state should have following
characters. Its mass, in principle, should range from $m_p +
m_{\phi}$ to $m_{J/\psi} -m_{\bar{p}}$, namely from 1.96GeV to
2.15GeV due to the phase space restriction. However, some $N^*$
states whose mass is smaller than the $p\phi$ threshold may also
contribute, because of their relatively larger branching fractions
for some decay channels involving strange particles and their
off-shell effect as well. The spin of the intermediate $N^*$ can
be any half-integer due to a relative angular momentum between
$N^*$ and $\bar{p}$. For minimizing our calculation without
affecting our qualitative conclusion, in the present approach, we
only consider those intermediate $N^*$ states whose contributions
may dominate the decay width. Thus, the selected $N^*$ state
should have following features: It should have a relatively large
branching fraction for a decay in which strange particles are
involved. As a consequence, it might have, according to the OZI
rule, a configuration with strangeness, so that it would be easier
decaying into $N\phi$ and relatively important in the $J/\psi \to
p \bar{p}\phi$ decay. It would also be reasonable to take the
embedded $N^*$ state with its spin up to $5/2$ only, since the
contribution from a $N^*$ state with higher spin would encounter a
power suppression due to a large relative angular momentum. In the
practical calculation, in the mass region above the $p\phi$
threshold, we only take $N^*_{1/2^-}(2090)S_{11}$ and
$N^*_{1/2^+}(2100)P_{11}$ with the $N\eta$ decay branching
fractions of about $0.41$ and $0.61$, respectively, although the
contribution from the later one would subject to a $p-$wave
suppression, and ignore $N^*_{3/2^-}(2080)D_{13}$ and
$N^*_{5/2^+}(2000)F_{15}$ because of their tiny $N\eta$ branching
fractions (about $0.03$). The situation in the sub-threshold
region is somewhat complex. We only take $N^*_{1/2^-}(1535)S_{11}$
and $N^*_{3/2^+}(1900)P_{13}$ into account due to their large
$N\eta$ branching fractions of about $0.45-0.60$ and $0.14$,
respectively. The reason for disregarding other sub-threshold
resonances like $N^*_{1/2^-}(1650)S_{11}$,
$N^*_{1/2^+}(1710)P_{11}$, $N^*_{3/2^+}(1720)P_{13}$,
$N^*_{5/2^+}(1680)F_{15}$, $N^*_{5/2^-}(1675)D_{15}$,
$N^*_{3/2^-}(1520)D_{13}$, $N^*_{1/2^+}(1440)P_{11}$, and
$N^*_{1/2^+}(939)P_{11}$ is as follows. For
$N^*_{1/2^-}(1650)S_{11}$, its branching fractions to $N\eta$ and
$\Lambda K$ are about 0.03$\sim$0.10 and 0.03$\sim$0.11,
respectively, which are much smaller than those for
$N^*_{1/2^-}(1535)S_{11}$. Moreover, one argued that due to the
weak coupling of $N^*_{1/2^-}(1650)S_{11}$ to $N\rho$ from SU(3)
symmetry, the coupling between $N^*_{1/2^-}(1650)S_{11}$ to
$N\phi$ might also be weak~\cite{j.j.xie}. In fact, if both
$N^*_{1/2^-}(1535)S_{11}$ and $N^*_{1/2^-}(1650)S_{11}$ are used
to fit the $\pi^{-}p\rightarrow n\phi$ data, the later one would
give an inappropriate contribution at the higher energies and the
fitted result shows an almost zero contribution from
$N^*_{1/2^-}(1650)S_{11}$~\cite{j.j.xie}. For
$N^*_{3/2^+}(1720)P_{13}$, its branching fractions to $N\eta$ and
$\Lambda K$ are less than 0.04 and about 0.01$\sim$0.15,
respectively, which are 2$\sim$3 times less than those for the
$N^*_{3/2^+}(1900)P_{13}$ state. For $N^*_{1/2^+}(1710)P_{11}$,
its branching fraction to $N\eta$ is less than 0.06, which is
almost 10 times less than that for $N^*_{1/2^+}(2100)P_{11}$, and
its branching fraction to $\Lambda K$ is about 0.05$\sim$0.25,
whose largest value is about the same as that for
$N^*_{1/2^+}(2100)P_{11}$. Additionally considering the factor
that $N^*_{3/2^+}(1720)P_{13}$ and $N^*_{1/2^+}(1710)P_{11}$ are
the states below the $N\phi$ threshold, their contributions would
be smaller than that from $N^*_{3/2^+}(1900)P_{13}$ and much
smaller than that from $N^*_{1/2^+}(2100)P_{11}$, respectively. In
fact, their contributions are evidently small near the $N\phi$
threshold region in the $\pi^{-} p \to n\phi$ reaction. For
$N^*_{1/2^+}(939)P_{11}$, $N^*_{1/2^+}(1440)P_{11}$,
$N^*_{3/2^-}(1520)D_{13}$, $N^*_{5/2^-}(1675)D_{15}$, and
$N^*_{5/2^+}(1680)F_{15}$, their branching fractions to the
$N\eta$ channel are almost zero, and the contributions from the
D-wave and F-wave states even suffer from the high-partial-wave
suppression~\cite{j.j.xie}. Based on such a discussion and the
result given by the partial wave analysis (PWA) in
Ref.~\cite{J.Z.Bai,M.Ablikim0}, we can safely assume that if
$N^*_{1/2^-}(1535)S_{11}$, $N^*_{3/2^+}(1900)P_{13}$,
$N^*_{1/2^-}(2090)S_{11}$ and $N^*_{1/2^+}(2100)P_{11}$ can give a
contribution about 85$\%$ to 90$\%$ of the total, taking these
four $N^*$ states to study a system with strange particles, for
instance the $J/\psi\rightarrow p\bar{p}\phi$ process, would be
meaningful. For simplicity, we omit the spectroscopic symbol in
the notation of the $N^*$ state hereafter.

To reveal the decay property of the $J/\psi\rightarrow
p\bar{p}\phi$ process, the coupling constants $g_{\phi N N^{*}}$ and
$g_{\psi N N^{*}}$ should be fixed at the beginning.

\subsection{Determination of $g^2_{\phi NN^{*}}$}

\begin{figure}[htbp]
\begin{center}
\includegraphics[scale=0.8]{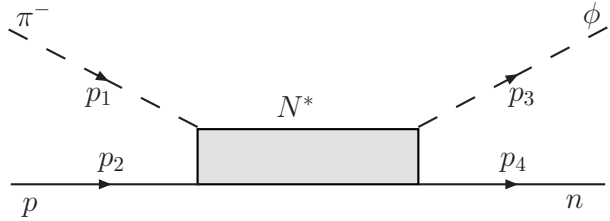} \caption{$s$-channel Feynman
diagram for the $\pi^- p \to n \phi$ reaction in the resonance
model.} \label{pipsu}
\end{center}
\end{figure}

As mentioned in Ref.~\cite{j.j.xie}, the $\phi$-meson production
near the threshold in the $\pi^{-} p \to n\phi$ reaction is
dominated by the intermediate nucleon resonances in the $s$-channel,
and the $u$-channel $N^*$ exchange and the $t$-channel $\rho$-meson
exchange between pion and proton are found to be negligible,
although in some references the $t$-channel $\rho$-meson exchange
and/or nucleon pole contributions were assumed to be
important~\cite{j.j.xie,SibHaidMeis}. Based on this argument, the
coupling constant $g^2_{\phi N N^{*}}$ can be extracted by fitting
the cross section data of the $\pi^{-} p \to n\phi$
reaction~\cite{pipdata,j.j.xie}. The $s$-channel Feynman diagram for
such a process is shown in Fig.~\ref{pipsu}, where $p_1, p_2, p_3,
p_4$, and $q$ denote the four-momenta of the incoming $\pi^-$ and
proton, outgoing $\phi$ and neutron, and intermediate $N^{*}$,
respectively.
In this diagram, the coupling constant $g^2_{\pi NN^*}$
($g^2_{\eta NN^*}$) can be determined in terms of a commonly used
effective Lagrangian~\cite{K.Tsushima,A.I.Titov,j.j.xie}. For a
nucleon resonance with $J^P_{N^*}=\frac{1}{2}^{-}$ where $J$ and
$P$ denote its spin and parity, respectively, the effective
Lagrangian can be written as~\cite{K.Tsushima,A.I.Titov,j.j.xie}
\begin{eqnarray}
{\cal L}_{\pi NN^*} & = & g_{\pi NN^*} \bar{N^*} \vec \tau \cdot
\vec \pi N + h.c., \label{pinnstaronehalf1}
\end{eqnarray}
with $J^P_{N^*}=\frac{1}{2}^{+}$
\begin{eqnarray}
{\cal L}_{\pi NN^*} & = & i g_{\pi NN^*} \bar{N^*} \gamma_5 \vec
\tau \cdot \vec \pi N + h.c., \label{pinnstaronehalf2}
\end{eqnarray}
and with $J^P_{N^*}=\frac{3}{2}^+$, say
$N^*(1900)$,~\cite{K.Tsushima}
\begin{eqnarray}
{\cal L}_{\pi NN^*} & = & i \frac{g_{\pi NN^*}}{M_{N^*}}
\bar{N}^{*\mu} \vec \tau \cdot \partial _{\mu}\vec \pi N + h.c.,
\label{pinnstarthreehalf}
\end{eqnarray}
where $g_{\pi NN^*}$, $N^{*\mu}$, $N$, $\vec{\pi}$ and $\vec{\tau}$
denote the coupling constant of a pion to a nucleon and a $N^*$, the
Rarita-Schwinger field of $N^*$ with its spin of 3/2 and mass of
$M_{N^*}$, the field of nucleon, the field of $\vec{\pi}$ and the
isospin matrices, respectively. And the effective Lagrangian for the
$\eta NN^{*}(1535)$ coupling can be expressed by:
\begin{eqnarray}
\mathcal{L}_{\eta N N^{*}}= g_{\eta N N^{*}}\bar{N^{*}}\eta N+h.c..
\end{eqnarray}

With these Lagrangians, the partial decay widths of the $N^*$
states can easily be derived by evaluating the transition from the
initial $N^*$ state to the final $N\pi~(N\eta)$ state
\begin{eqnarray}
\Gamma_{N^*(1535) \to N \pi} &=& \frac{3 g^2_{\pi
NN^*}(m_N+E_N)p^{cm}_{\pi}}{4\pi M_{N^*}},
 \label{eq:1535pi}
\end{eqnarray}
\begin{eqnarray}
\Gamma_{N^*(1900) \to N \pi} &=& \frac{g^2_{\pi
NN^*}(m_N+E_N)(p^{cm}_{\pi})^3}{4\pi M^3_{N^*}},
 \label{eq:1900pi}
\end{eqnarray}
\begin{eqnarray}
\Gamma_{N^*(2090) \to N \pi} &=& \frac{3 g^2_{\pi
NN^*}(m_N+E_N)p^{cm}_{\pi}}{4\pi M_{N^*}},
 \label{eq:2090pi}
\end{eqnarray}
\begin{eqnarray}
\Gamma_{N^*(2100) \to N \pi} &=& \frac{3 g^2_{\pi
NN^*}(E_N-m_N)p^{cm}_{\pi}}{4\pi M_{N^*}},
 \label{eq:2100pi}
\end{eqnarray}
\begin{eqnarray}
\Gamma_{N^*(1535) \to N \eta} &=& \frac{g^2_{\eta
NN^*}(m_N+E_N)p^{cm}_{\eta}}{4\pi M_{N^*}}
 \label{eq:1535eta}
\end{eqnarray}
with
\begin{equation}
p^{cm}_{\pi(\eta)}=\sqrt{\frac{(M^2_{N^*}-(m_N+m_{\pi(\eta)})^2)
(M^2_{N^*}-(m_N-m_{\pi(\eta)})^2)}{4M^2_{N^*}}},
\end{equation}
and
\begin{equation}
E_N=\sqrt{(p^{cm}_{\pi(\eta)})^2+m^2_N}.
\end{equation}
By re-producing the mass, the width and the $\pi N$($\eta N$)
channel branching fraction of the $N^*$ state~\cite{pdg2008}
measured in the experiment, the phenomenological coupling constant
$g^2_{\pi N N^*}$ ($g^2_{\eta N N^*}$) can be extracted.

Furthermore, the effective Lagrangians of the $\phi NN^*$
interaction for various $N^*$ states are adopted as follows: For a
$N^*$ with $J^P_{N^*}=\frac{1}{2}^-$, say $N^*(1535)$ or
$N^*(2090)$, ~\cite{j.j.xie}
\begin{eqnarray}
{\cal L}_{\phi NN^*} & = &  g_{\phi NN^*} \bar{N^*} \gamma_5
(\gamma_{\mu} -\frac{q_{\mu}\not\! q}{q^2}) \Phi^{\mu} N + h.c.
\label{phiNNstar1535}
\end{eqnarray}
with $q$ being the four-momentum of $N^{*}$ and $\Phi^{\mu}$ being
the field of $\phi$ meson. For a $N^*$ with
$J^P_{N^*}=\frac{1}{2}^+$, say $N^*(2100)$,
\begin{eqnarray}
{\cal L}_{\phi NN^*} & = &  g_{\phi NN^*} \bar{N^*}
\gamma_{\mu}\Phi^{\mu}_{\phi} N + h.c., \label{phiNNstar2100}
\end{eqnarray}
and for a $N^*$ with $J^P_{N^*}=\frac{3}{2}^+$, say
$N^*(1900)$,~\cite{K.Tsushima}
\begin{eqnarray}
{\cal L}_{\phi NN^*} & = & i g_{\phi NN^*} \bar{N}^{*}_{\mu}
\gamma_5 \Phi^{\mu}_{\phi} N + h.c.. \label{phiNNstar1900}
\end{eqnarray}
To reckon for the off-shell effect of $N^*$, a form factor
\begin{equation}
F_{N^*}(q^2)=\frac{\Lambda^{4}}{\Lambda^{4} + (q^2-M^2_{N^*})^2}
\label{ffNstar}
\end{equation}
with $\Lambda$ being the cut-off parameter is introduced in the
$MNN^*$ vertex~\cite{Mosel, feuster}.

The propagator $G_{J^P}(q)$ of a $N^*_{J^P}$ with the quantum
number $J^P$ and momentum $q$ can be written in a Breit-Wigner
form~\cite{liang}. For the $J_{N^*}={1/2}$ state,
\begin{equation}
G_{\frac{1}{2}^{\pm}}(q)=\frac{i(\pm\not\! q
+M_{N^*})}{q^2-M^2_{N^*}+iM_{N^*}\Gamma_{N^*}},
\end{equation}
where $\Gamma_{N^*}$ denotes the total decay width of the $N^*$
state, and the $+$ and $-$ signs on the left of $\not\! q$ are the
signs for the positive and negative parity states, respectively.
For the $J_{N^*}={3/2}$ state,
\begin{equation}
G^{\mu \nu}_{\frac{3}{2}^{\pm}}(q)=\frac{i(\pm\not\! q
+M_{N^*})}{q^2-M^2_{N^*}+iM_{N^*}\Gamma_{N^*}} (-g^{\mu \nu}
+\frac{1}{3}\gamma_{\mu}\gamma_{\nu}\mp\frac{1}{3M_{N^*}}
(\gamma^{\mu}q^{\nu}-\gamma^{\nu}q^{\mu})+\frac{2}{3M_{N^*}}
q^{\mu}q^{\nu}).
\end{equation}

In terms of the effective Lagrangian, form factor and $N^*_{J^P}$
propagator mentioned above, we use Feynman rules to write the
invariant amplitude contributed by a $N^*$ in the $s-$channel
$\pi^{-} p \to n\phi$ reaction as
\begin{eqnarray}
\label{eq:pimptonphi} {\cal M}_{N^*}^{\pi^{-}p}\propto &&
\sqrt{2}g_{\pi N N^{*}}g_{\phi N
N^{*}}F_{N^{*}}(q^{2})\bar{u}(p_{n},s_{n})\Gamma_{\phi NN^*}
\varphi_{\phi}(p_{\phi},s_{\phi})\nonumber \\
&&G_{N^{*}}(q)\varphi_{\pi}(p_{\pi},s_{\pi})\Gamma_{\pi
NN^*}u(p_{p},s_{p}),
\end{eqnarray}
where $u$, $\varphi_{\phi}$ and $\varphi_{\pi}$ denote the fields
of the nucleon, $\phi-$meson and $\pi-$meson, respectively, $p_n$,
$p_p$, $p_{\phi}$ and $p_{\pi}$ represent the momenta of the
proton, neutron, $\phi-$meson and $\pi-$meson, respectively,
$s_n$, $s_p$, $s_{\phi}$ and  $s_{\pi}$ describe the spins of the
proton, neutron, $\phi-$meson and $\pi-$meson, respectively, and
$\Gamma_{\phi NN^*}$ and $\Gamma_{\pi NN^*}$ stand for the vertex
functions of $\phi NN^*$ and $\pi NN^*$, respectively. The
formulae of the invariant amplitude for various $N^*$s are given
in Appendix A. Summing up all amplitudes for the $N^*$s
considered, we obtain the total invariant amplitude
\begin{eqnarray}
{\cal M}_{\pi^{-} p\rightarrow n\phi}=\sum_{N^*}{\cal
M}_{N^*}^{\pi^{-}p},
\end{eqnarray}
where $N^*$ runs over all the considered states. Consequently, we
can calculate the total cross section of the $\pi^{-} p \to n\phi$
reaction by using the following equation
\begin{eqnarray}
\sigma=\int
d\Phi_2(\mathbb{P},p_p,p_{\pi},p_n,p_{\phi})\frac{(2\pi)^4}{4\sqrt{(p_p
\cdot p_{\pi})^2-m_{p}^2m_{\pi}^2}}|{\cal M}_{\pi^{-} p\rightarrow
n\phi}|^2,
\end{eqnarray}
with $d\Phi_2$ being an element of the two-body phase space, and
$\mathbb{P}$ being the total momentum of the system. By adjusting
the coupling constants $g^2_{\phi NN^*}$ to fit the total cross
section of the $\pi^{-} p\rightarrow n\phi$ reaction, we can
extract a set of phenomenological $g^2_{\phi NN^*}$. We would
further mention that the contributions from the $u-$channel and
meson-exchange channel will not be included in the calculation,
because they are negligibly small~\cite{j.j.xie}.

\subsection{Determination of $g^2_{\psi NN^{*}}$}

The coupling constant $g^2_{\psi NN^*}$ can be extracted from the
BESII data for the $J/\psi\rightarrow p \bar{n}\pi^{-}$ and
$J/\psi\rightarrow p \bar{p}\eta$
decays~\cite{J.Z.Bai,M.Ablikim0}. The Feynman diagrams for these
decays are the same as those in Fig.\ref{ppbarphi} except that the
$\phi$-meson is replaced with the $\pi$- and $\eta$-mesons,
respectively.

The effective Lagrangian for the $J/\psi NN^*$ interaction can be
chosen in the following form: For a $N^*$ with
$J^P_{N^*}=\frac{1}{2}^-$, say $N^*(1535)$ or
$N^*(2090)$,~\cite{b.ch.liu1}
\begin{eqnarray}
\mathcal{L}_{\psi N N^{*}}=ig_{\psi N
N^{*}}\bar{N^{*}}\gamma_{5}\sigma_{\mu\nu}p^{\nu}_{\psi}
\epsilon^{\mu}(\vec{p}_{\psi},s_{\psi}) N+h.c.
\end{eqnarray}
with $p_{\psi}$ and $\varepsilon(p_{\psi})$ being the four-momentum
and the polarization vector of $J/\psi$, respectively, for a $N^*$
with $J^P_{N^*}=\frac{1}{2}^+$, say $N^*(2100)$,
\begin{eqnarray}
 {\cal L}_{\psi NN^*} & = &
g_{\psi NN^*} \bar{N^*}
\gamma_{\mu}\epsilon^{\mu}(\vec{p}_{\psi},s_{\psi}) N + h.c.,
\label{phiNNstar2100}
\end{eqnarray}
and for a $N^*$ with $J^P_{N^*}=\frac{3}{2}^+$, say
$N^*(1900)$,~\cite{K.Tsushima}
\begin{eqnarray}
{\cal L}_{\psi NN^*} & = & i g_{\psi NN^*} \bar{N}^{*}_{\mu}
\gamma_5 \epsilon^{\mu}(\vec{p}_{\psi},s_{\psi}) N + h.c..
\label{phiNNstar1900}
\end{eqnarray}

It should be mentioned that $J/\psi$ meson produced in BEPCII is
transversely polarized, namely $s_{3}=\pm1$. The completeness
condition of polarization vector obeys 
\begin{eqnarray}
\sum_{s=\pm1}\epsilon_{\mu}(\vec{p},s)
\epsilon^{*}_{\nu}(\vec{p},s)=
\delta_{\mu\nu}(\delta_{\mu1}+\delta_{\mu2}),
\end{eqnarray}
where $\epsilon_{\mu}(\vec{p},s)$, $\vec{p}$ and $s$ denote the
polarization vector, the momentum, and the polarization direction of
$J/\psi$, respectively, and $\delta$ is a Kronecker Delta symbol.

Then, the invariant decay amplitude contributed by a specific $N^*$
in the $J/\psi\rightarrow p \bar{n}\pi^{-}$ ($J/\psi\rightarrow p
\bar{p}\eta$) decay can easily be written as
\begin{eqnarray}
\label{eq:deacy} {\cal M}_{N^*}^{J/\psi~ decay}\propto && \xi
g_{\pi(\eta) N N^{*}}g_{\psi
NN^{*}}F_{N^{*}}(q^{2})\bar{u}(p_{p},s_{p})\Gamma_{\pi(\eta)NN^*}
\varphi_{\pi(\eta)}(p_{\pi(\eta)},s_{\pi(\eta)})\nonumber \\
&&G_{N^{*}}(q) \varphi_{\psi}(p_{\psi},s_{\psi})\Gamma_{\psi NN^*}
v(p_{\bar{p}},s_{\bar{p}})
\end{eqnarray}
with $u$ ($v$) being the field of proton (anti-proton),
$\varphi_{\psi}$ and $\varphi_{\pi(\eta)}$ being the fields of
$\psi$ and $\pi(\eta)$, respectively, $p_{p}$, $p_{\bar{p}}$,
$p_{\pi(\eta)}$ and $p_{\psi}$ being the momenta of the proton,
anti-proton, $\psi$-meson and $\pi$- ($\eta$-)meson, respectively,
$s_{p}$, $s_{\bar{p}}$, $s_{\pi(\eta)}$ and $s_{\psi}$ being the
spins of the proton, anti-proton, $\psi$-meson and $\pi$-
($\eta$-)meson, respectively, and $\Gamma_{\pi(\eta)NN^*}$ and
$\Gamma_{\psi NN^*}$ being the vertex functions of $\pi(\eta)NN^*$
and $\psi NN^*$, respectively. The coefficient $\xi$ is taken to
be $\sqrt{2}$ for the $p\bar{n}\pi^{-}$ reaction, but 1 for the
$p\bar{p}\eta$ reaction. The formulae of the invariant amplitude
for various $N^*$s are given in Appendix B. Consequently, the
total invariant amplitude can be obtained by summing over all
possible $N^*$ states
\begin{eqnarray}
{\cal M}_{J/\psi~decay}=\frac{1}{2}\sum_{N^*}{\cal
M}_{N^*}^{J/\psi~decay},
\end{eqnarray}
where the factor of 1/2 comes from the average over the $J/\psi$
spin. The partial decay width of $J/\psi \to p \bar{n} \pi^{-}$
($J/\psi \to p \bar{p} \eta$) can be calculated by
\begin{eqnarray}
d\Gamma= &&
\frac{1}{2}\frac{1}{2M_{\psi}}\frac{p^{0}_{p}d^{3}p_{p}}{m_{N}}
\frac{p^{0}_{\bar{p}}d^{3}p_{\bar{p}}}{m_{N}}
\frac{d^{3}p_{\pi(\eta)}}{2p^{0}_{\pi(\eta)}}\nonumber \\
&& \sum_{s_{\psi}}\sum_{s_{p},s_{\bar{p}},s_{\pi(\eta)}} |{\cal
M}_{J/\psi~
decay}|^{2}(2\pi)^{-5}\delta^{4}(p_{\psi}-p_{p}-p_{\bar{p}}-p_{\pi(\eta)}).
\end{eqnarray}
By fitting the branching fractions of ($2.09\pm 0.18) \times
10^{-3}$ for $J/\psi\rightarrow p \bar{n}\pi^{-}$ and ($2.12\pm
0.09) \times 10^{-3}$ for $J/\psi\rightarrow p
\bar{p}\eta$~\cite{pdg2008}, respectively, the magnitude of
$g^2_{\psi N N^{*}}$ can be extracted.

\subsection {$J/\psi\rightarrow p\bar{p}\phi$ decay}
Using the effective Lagrangians mentioned above, the invariant
amplitude of the $J/\psi\rightarrow p\bar{p}\phi$ decay can easily
be derived. Its form is the same as that in Eq.(\ref{eq:deacy})
except that $\pi$ is substituted with $\phi$
\begin{eqnarray}
\label{eq:deacy-phi} {\cal M}_{N^*}^{J/\psi \to p
\bar{p}\phi}\propto && g_{\phi N N^{*}}g_{\psi
NN^{*}}F_{N^{*}}(q^{2})\bar{u}(p_{p},s_{p})\Gamma_{\phi NN^*}
\varphi_{\phi}(p_{\phi},s_{\phi})\nonumber \\
&&G_{N^{*}}(q) \varphi_{\psi}(p_{\psi},s_{\psi})\Gamma_{\psi NN^*}
v(p_{\bar{p}},s_{\bar{p}}).
\end{eqnarray}
The formulae of the invariant amplitude contributed by various
$N^*$s are given in Appendix C. Then, the total invariant
amplitude can be obtained by summing over the contributions from
all possible $N^*$ states
\begin{eqnarray}
{\cal M}_{J/\psi \to p\bar{p}\phi}=\frac{1}{2}\sum_{N^*}{\cal
M}^{J/\psi \to p\bar{p}\phi}_{N^*},
\end{eqnarray}
where the factor of 1/2 is due to the average over the $J/\psi$
spin as usual. The invariant mass spectrum of $p\phi$ in the
$J/\psi \to p\bar{p}\phi$ decay can be expressed as~\cite{pdg2008}
\begin{eqnarray}
\frac{d\Gamma}{d\Omega^{*}_{p}d\Omega_{\bar{p}}}=\frac{1}{2}
\frac{1}{(2\pi)^{5}}\frac{(2m_{p})^{2}}{16M^{2}_{\psi}}{|{\cal
M}_{J/\psi \to p \bar{p}\phi}|^{2}}|p^{*}_{p}||\bar{p}|dm_{p\phi},
\end{eqnarray}
where ($|p^{*}_{p}|, \Omega^{*}_{p}$) is the momentum of proton in
the rest frame of $p$ and $\phi$, and $d\Omega_{\bar{p}}$ is the
angle of anti-proton in the rest frame of the decaying $J/\psi$.
Integrating over all the angles in the rest frame of $J/\psi$, the
Dalitz plot can be derived in the following form~\cite{pdg2008}
\begin{eqnarray}
d\Gamma=\frac{1}{2}\frac{1}{(2\pi)^{3}}\frac{(2m_{p})^{2}}{32M_{\psi}^3}{|{\cal
M}_{J/\psi \to p \bar{p}\phi}|^{2}}
dm_{p\phi}^{2}dm_{\bar{p}\phi}^{2},
\end{eqnarray}
with
\begin{eqnarray}
m_{p\phi}^{2}=p_{p\phi}^{2}=(p_{\psi}-p_{\bar{p}})^2,
~~~~~m_{\bar{p}\phi}^{2}=p_{\bar{p}\phi}^{2}=(p_{\psi}-p_{p})^2.
\end{eqnarray}

\section{RESULTS AND DISCUSSION}
Based on the discussion in the last section, only
$N^*_{1/2^-}(1535)$, $N^*_{3/2^+}(1900)$, $N^*_{1/2^-}(2090)$ and
$N^*(2100)P_{11}$ are adopted as the intermediate state in the
practical calculation. The coupling constant $g_{\pi(\eta)NN^*}$
for these $N^*$s are extracted by using the decay width formulas
for $N^* \to N\pi(\eta)$ shown in the last section, where the
masses of $N$, $N^*$, $\pi$ and $\eta$ are taken from
PDG~\cite{pdg2008}, namely $m_N$=0.938GeV, $m_{\pi}$=0.139GeV and
$m_{\eta}$=0.547GeV, $M_{N^*(1535)}$=1.535GeV,
$M_{N^*(1900)}$=1.900GeV, $M_{N^*(2090)}$=2.090GeV, and
$M_{N^*(2100)}$=2.100GeV, respectively~\cite{pdg2008}.

It should be noted that the total width and the branching
fractions of $N^*_{S_{11}}(1535)$ (or the partial decay widths)
for the $N\pi$ and $N\eta$ channels have more or less accurately
been measured, thus $g^2_{\pi NN^*(1535)}$ and $g^2_{\eta
NN^*(1535)}$ can be estimated by using the averaged values of
branching fractions given in PDG~\cite{pdg2008}. However, for the
two-star state $N^*_{3/2^+}(1900)$ and one-star states
$N^*_{1/2^-}(2090)$ and $N^*_{1/2^+}(2100)$, their partial decay
widths for the $N\pi$ channel have not precisely been confirmed
yet. The extracted $g^2_{\pi NN^*}$ would be allowed to change in
a range due to the mentioned large uncertainty. The range can
roughly be estimated by using the maximal and minimal
values~\cite{pdg2008} of the total width and the $N\pi$ branching
fraction for the corresponding $N^*$. The extracted $g^2_{\pi
NN^{*}}$ and $g^2_{\eta NN^{*}}$ for each $N^*$ are tabulated in
Table~\ref{tab:nstarpip}.
\begin{table}[htbp]
\centering \caption{Coupling constants $g^2_{\pi NN^{*}}$ and
$g^2_{\eta NN^{*}}$for various $N^*$ states.}
\begin{tabular}{|c|c|c|c|c|c|}
\hline \multirow{2}{*}{$N$}           & total width  &
\multirow{2}{*}{decay mode}  & branching fraction  & partial width
& \multirow{2}{*}{$g^2_{\pi NN^{*}}$
($g^2_{\eta NN^{*}}$)} \\ 
 & (GeV)\cite{pdg2008} &  & \cite{pdg2008}   & (GeV) &
 \\  \hline
\multirow{2}{*}{$N^{*}_{1/2^-}(1535)$} & 0.150  & $N\pi$ & $45\%$
& 0.675$\times 10^{-1}$ &  0.468 \\
\cline{2-6}
 & 0.150 & $N\eta$ & $53\%$
 & 0.795$\times 10^{-1}$ &  0.431$\times 10^{1}$  \\
\hline %
\multirow{2}{*}{$N^{*}_{3/2^+}(1900)$} & 0.180 & $N\pi$ & $5.5\%$
& 0.990$\times 10^{-2}$ &  0.113$\times 10^{1}$ \\
\cline{2-6}
 & 0.498 & $N\pi$ & $26\%$
& 0.129 &  0.147$\times 10^{2}$ \\
\hline %
\multirow{3}{*}{$N^{*}_{1/2^-}(2090)$}  & 0.095  & $N\pi$  &
$9.0\%$
& 0.855$\times 10^{-2}$ & 0.410 $\times 10^{-1}$ \\
\cline{2-6}
 & 0.350  & $N\pi$   & $18\%$
& 0.630$\times 10^{-1}$ & 0.305   \\
\cline{2-6}
 & 0.414  & $N\pi$   & $10\%$
& 0.414$\times 10^{-1}$ & 0.200   \\
\hline %
\multirow{3}{*}{$N^{*}_{1/2^+}(2100)$}  & 0.113  & $N\pi$  &
$15\%$
& 0.170$\times 10^{-1}$ & 0.564 \\
\cline{2-6}
 & 0.200  & $N\pi$   & $10\%$
& 0.200$\times 10^{-1}$ & 0.666   \\
\cline{2-6}
 & 0.260  & $N\pi$   & $12\%$
& 0.312$\times 10^{-1}$ & 0.104$\times 10^{1}$   \\
\hline %
\end{tabular}
 \label{tab:nstarpip}
\end{table}
From this table, it is clearly shown that the result is reasonable,
namely it consists with the fact that the larger the partial decay
width is, the stronger the $N^*$ couples to the decayed particles.

Then the coupling constants $g_{\phi N N^{*}}$ for various $N^*$s
can be extracted by fitting the total cross section data for the
$\pi^{-}p\rightarrow n\phi$ reaction. To consider the off-shell
effect of $N^*$, a form factor in Eq.(\ref{ffNstar}) with a
cut-off parameter $\Lambda$ being 1.8GeV for $N^*_{1/2^-}(1535)$
and 2.3GeV for $N^*_{3/2^+}(1900)$, $N^*_{1/2^-}(2090)$ and
$N^*_{1/2^+}(2100)$ is employed.

\begin{figure}[htbp]
\centering \subfigure[Type
I($\Gamma_{N^{*}(1900)}$/$\Gamma_{N^{*}(2090)}$/$\Gamma_{N^{*}(2100)}$=
498MeV/95MeV/260MeV)]{
\label{fig:subfig:d} 
\includegraphics[width=8cm]{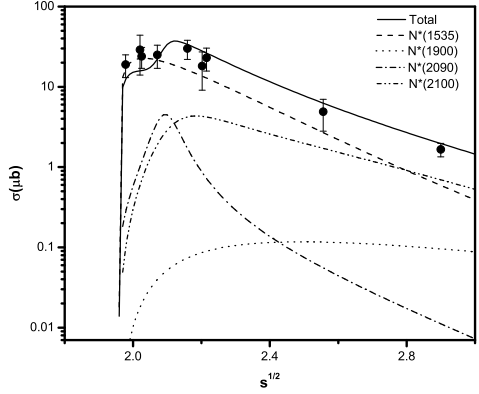}}
\subfigure[Type
II($\Gamma_{N^{*}(1900)}$/$\Gamma_{N^{*}(2090)}$/$\Gamma_{N^{*}(2100)}$=
180MeV/350MeV/113MeV)]{
\label{fig:subfig:b} 
\includegraphics[width=8cm]{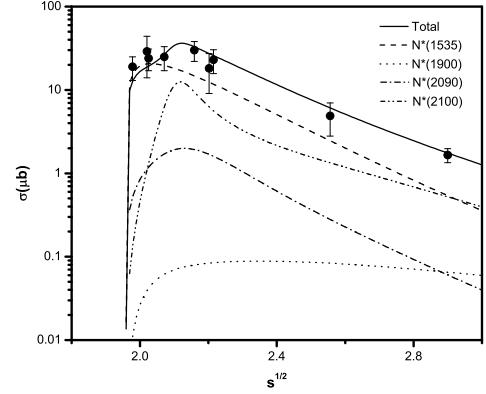}}
\subfigure[Type
I($\Gamma_{N^{*}(1900)}$/$\Gamma_{N^{*}(2090)}$/$\Gamma_{N^{*}(2100)}$=
498MeV/95MeV/260MeV)]{
\label{fig:subfig:f} 
\includegraphics[width=8cm]{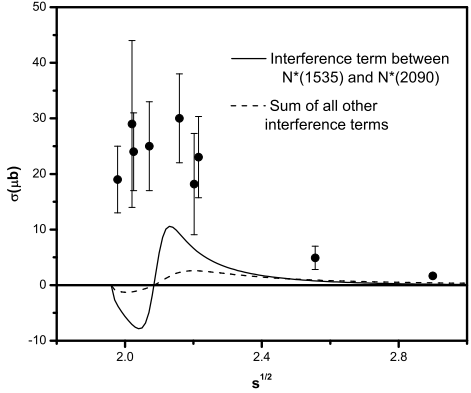}}
\subfigure[Type
II($\Gamma_{N^{*}(1900)}$/$\Gamma_{N^{*}(2090)}$/$\Gamma_{N^{*}(2100)}$=
180MeV/350MeV/113MeV)]{
\label{fig:subfig:e} 
\includegraphics[width=8cm]{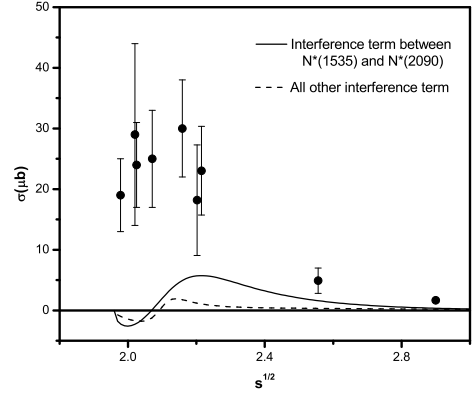}}
\caption{Total cross section of the $\pi^{-}p\rightarrow n\phi$
reaction.}
\label{pimp} 
\end{figure}

Because the magnitudes of $g^2_{\pi NN^{*}}$ for the
$N^*_{3/2^+}(1900)$, $N^*_{1/2^-}(2090)$ and $N^*_{1/2^+}(2100)$
states can respectively vary in a rather large range of their own,
we combine possible $g^2_{\pi NN^{*}}$ values for these $N^*$s
into various cases to fit the reaction data. The fitted result
shows that only two types of combinations can give the best fit.
In type I, one can only take a smaller total width for
$N^*_{1/2^-}(2090)$ and a larger total width for
$N^*_{1/2^+}(2100)$, and in type II, it is the other way round. To
be specific, the restricted regions of the total widths for
$N^*_{3/2^+}(1900)$, $N^*_{1/2^-}(2090)$ and $N^*_{1/2^+}(2100)$
are bound by the following combined cases:
$\Gamma_{N^{*}(1900)}$/$\Gamma_{N^{*}(2090)}$/$\Gamma_{N^{*}(2100)}$
=180MeV/95MeV/200MeV, 180MeV/95MeV/260MeV, 498MeV/95MeV/200MeV,
and 498MeV/95MeV/260MeV in type I, and 180MeV/350MeV/113MeV,
180MeV/414MeV/113MeV, 498MeV/350MeV/113MeV and
498MeV/414MeV/113MeV in type II. The typical fitted curves of
these two types are plotted in Figs.~\ref{pimp} (a) and (b) with
$\chi^2$ being $3.62$ and $3.05$, respectively. In these figures,
the dashed, dotted, dash-dotted and dash-double-dotted curves
denote the contributions from individual $N^*_{1/2^-}(1535)$,
$N^*_{3/2^+}(1900)$, $N^*_{1/2^-}(2090)$ and $N^*_{1/2^+}(2100)$
states, respectively. The contributions from interference terms in
these cases are shown in Figs.~\ref{pimp} (c) and (d),
respectively. In these figures, we only plot the the contributions
from the interference term between $N^{*}_{1/2^-}(1535)$ and
$N^{*}_{1/2^-}(2090)$ and from the sum of the rest terms, shown as
the solid and dashed curves, respectively, because the former is
much larger than the later. The fitted cures are plotted by solid
curves in Figs.~\ref{pimp} (a) and (b), respectively. They are
obtained by summing over the contributions from all the $N^*$
states coherently.

From Fig.~\ref{pimp}, we find that $N^*_{1/2^-}(1535)$ provides a
major contribution in the whole energy range considered,
especially near the $N\phi$ threshold. The contribution from
$N^*_{3/2^+}(1900)$ is relatively flat in the high energy region.
The cross section from $N^*_{1/2^-}(2090)$ or $N^*_{1/2^+}(2100)$
has large uncertainty. Its shape depends on the total width of the
state, $\Gamma_{N^*}$. If $\Gamma_{N^*}$ is small, the cross
section curve would show a relatively narrow peak around the mass
of the $N^*$, otherwise it presents a broad structure. This is
simply because that a Breit-Wigner form for the $N^*$ propagator
is adopted in the calculation. The $P$ wave $N^*_{1/2^+}(2100)$
state can only play a minor role although it has a large branching
fraction to $N\eta$, since its contribution near the $N\phi$
threshold is too small. Furthermore, the contribution from
$N^*(2090)$ cannot be large because of a counter-contribution from
the interference term between $N^*(2090)$ and $N^*(1535)$ in a
region close to the $N\phi$ threshold, namely a larger
contribution from $N^*(2090)$ makes the fit worse. In conclusion,
although some higher resonances are introduced, the dominate
contribution in the $\pi^{-}p\rightarrow n\phi$ cross section
still comes from the $N^*_{1/2^-}(1535)$ state, which is
consistent with discussion in Ref.~\cite{j.j.xie}. It should
further be mentioned that the contribution from all the
interference terms is about 2\% only. Thus, the assumption that
the contribution from ignored $N^*$s including their interference
terms is about 10$\sim$ 15\% of the total would be reasonable, and
arranging the contributions from mentioned four $N^*$s in a range
of 85$\sim$ 90\% of the total will not affect our qualitative
conclusion.

Based on the best fit, namely a small enough $\chi^2$ and a
reasonable overall fit, we can extract the coupling constant
$g^{2}_{N^{*}N\phi}$ for all adopted $N^*$s in all the mentioned
cases. The resultant $g^{2}_{N^{*}N\phi}$ for these $N^*$s are
tabulated in Table~\ref{tab:pipnphi}. The fractions of the
individual contributions from all the considered $N^*$s are given
in the table as well.
\begin{table}[htbp]
\centering \caption{The extracted coupling constant
$g^{2}_{N^{*}N\phi}$ and the corresponding fraction of contribution
in the $\pi^- p \to n \phi$ reaction.}
 \label{tab:pipnphi}
\begin{tabular}{|c|c|c|c|c|c|} \hline
 & \multirow{2}{*}{$\Gamma_{N^{*}(1900)}$/ $\Gamma_{N^{*}(2090)}$/
$\Gamma_{N^{*}(2100)}$} &
\multicolumn{4}{|c|}{$g^{2}_{\phi NN^{*}}$($10^{-2}$)/(fraction of contribution($\%$))}  \\
\cline{3-6} & &   $N^{*}_{1/2^-}(1535)$& $N^{*}_{3/2^+}(1900)$ &
$N^{*}_{1/2^-}(2090)$ &$N^{*}_{1/2^+}(2100)$ \\\hline
\multirow{4}{*}{type I} & {180MeV/95MeV/200MeV} &140/67.1
&$7.92$/0.9 & 0.937/3.6 &$12.3$/13.5
\\\cline{2-6}
&{498MeV/95MeV/200MeV} & 137/64.9  &1.14/1.0 & 1.19/4.8 &9.46/9.8 \\ \cline{2-6} %
&{180MeV/95MeV/260MeV} & 128/61.7  &3.53/0.4 & 1.37/5.6 &10.8/15.4
\\ \cline{2-6}
&{498MeV/95MeV/260MeV}  & 126/60.0  &0.758/0.8 & 1.27/4.7 &11.7/16.4 \\ \hline %
\multirow{4}{*}{type II}&{180MeV/350MeV/113MeV} &116/55.2 &6.41/0.7
& 0.749/4.4 &16.2/23.3
\\\cline{2-6}
&{498MeV/350MeV/113MeV} & 115/52.6  &0.656/0.7 & 0.967/5.4 &17.0/23.1 \\ \cline{2-6} %
&{180MeV/414MeV/113MeV} & 118/56.2  &3.27/0.4 & 1.13/3.7 &17.5/24.8
\\ \cline{2-6}
&{498MeV/414MeV/113MeV}  & 114/54.0  &0.783/0.7 & 1.05/3.3 &18.6/25.7 \\ \hline %
\end{tabular}

\end{table}
From this table, one has following observations: (1)
$N^*_{1/2^-}(1535)$ is a dominant resonance in the
$\pi^{-}p\rightarrow n\phi$ reaction and provides about 50\% to 70\%
of the contribution. This state may couple to $N\phi$ strongly, and
the coupling constant $g^{2}_{N^{*}(1535)N\phi}$ ranges from 1.1 to
1.4. The contribution and the coupling to $N\phi$ in type I is
larger than those in type II. (2) The $N^*_{1/2^+}(2100)$ state is
the second largest contributor, which offers about 10\% to 26\% of
the contribution. It also shows that $N^*_{1/2^+}(2100)$ may couple
to $N\phi$ remarkably. The value of the coupling constant
$g^{2}_{N^{*}(2100)N\phi}$ stretches from 0.09 to 0.19. And the
contribution and the coupling to $N\phi$ in type II is larger than
those in type I. (3) The contribution from $N^*_{1/2^-}(2090)$ is
about 3\% to 6\%, and $g^{2}_{N^{*}(2090)N\phi}$ spans a range from
0.007 to 0.014. The contribution and the coupling to $N\phi$ in type
I is slightly larger than those in type II. (4) The contribution
from $N^*_{3/2^+}(1900)$ is even smaller, about 0.4 \% to 1.0\%, and
$g^{2}_{N^{*}(1900)N\phi}$ spreads in a range of 0.006 to 0.079. The
effect from the total width uncertainty of this state is quite
small. These observations are clearly consistent with the
information from the curves shown in Fig.~\ref{pimp}. Namely,
$N^*_{1/2^+}(2100)$ gives a contribution comparable to that from
$N^*_{1/2^-}(1535)$, especially in the higher energy region,
$N^*_{1/2^-}(2090)$ offers a visible contribution around the energy
about its mass, and $N^*_{3/2^+}(1900)$ only provides a very small
contribution in the whole energy region.

Next, we determine $g^2_{\psi N N^{*}}$ in terms of the partial
decay widths of the $J/\psi\rightarrow p\bar{p}\eta$ and
$J/\psi\rightarrow p\bar{n}\pi^{-}$ processes, respectively. The
partial wave analysis of the $J/\psi \to p \bar{p} \eta$ data
collected at BESII shows that the partial decay width contributed by
the intermediate $N^*_{1/2^-}(1535)$ state is about
$(56\pm15)\%$~\cite{J.Z.Bai}. By fitting this width, one can easily
obtain the value of $g^2_{\psi N N^{*}(1535)}$. Again, a form factor
with $\Lambda$ being 1.8GeV in Eq.(\ref{ffNstar}) is adopted in the
calculation to describe the off-shell effect of $N^*_{1/2^-}(1535)$,
and the extracted $g^2_{\psi N N^{*}(1535)}$ is tabulated in
Table~\ref{tab:nnpsi}. On the other hand, one notices that in
analyzing the $J/\psi \to p \bar{n} \pi^{-}$ data of BESII, assuming
the contributions from $N^*_{3/2^+}(1900)$, $N^*_{1/2^-}(2090)$ and
$N^*_{1/2^+}(2100)$ to be about 5$\sim$10\%, respectively, are
reasonable, and the resultant branching fraction of this channel is
about $(1.33\pm 0.02(stat.))\times 10^{-3}$~\cite{M.Ablikim0}.
Therefore, we also approximately take the contributions from
$N^*_{1/2^-}(1535)$, $N^*_{3/2^+}(1900)$, $N^*_{1/2^-}(2090)$ and
$N^*_{1/2^+}(2100)$ to be 56\%, 10\%, 10\% and 10\%, respectively,
in the calculation of the $J/\psi \to p \bar{n} \pi^{-}$ decay.
Using these assumptions, the extracted $g^2_{\pi N N^{*}(1535)}$
values and the form factor in Eq.(\ref{ffNstar}) with $\Lambda$
being 1.8GeV for $N^*_{1/2^-}(1535)$ and 2.3GeV for either
$N^*_{3/2^+}(1900)$, or $N^*_{1/2^-}(2090)$ or $N^*_{1/2^+}(2100)$,
we can extract $g^2_{\psi N N^{*}}$ for later three $N^*$s from the
the branching fraction of the $J/\psi \to p \bar{n} \pi^{-}$ decay.
The resultant $g^2_{\psi N N^{*}}$ and corresponding $\Lambda$ are
tabulated in Table~\ref{tab:nnpsi}.
\begin{table}[htbp]
\centering \caption{$g^2_{\psi NN^{*}}$ and $\Lambda$ for
$N^*_{1/2^-}(1535)$, $N^*_{3/2^+}(1900)$, $N^*_{1/2^-}(2090)$ and
$N^*_{1/2^+}(2100)$.}
\begin{tabular}{|c|c|c|c|}
\hline \multirow{2}{*}{$N^*$} & \multirow{2}{*}{Total Width(GeV)} &
\multicolumn{2}{|c|}{$g^{2}_{\psi N N^{*}}$} \\\cline{3-4}
 & & $\Lambda$=1.8GeV & $\Lambda$=2.3GeV \\ \hline
$N^{*}_{1/2^-}(1535)$ & 0.150  & 1.319$\times10^{-6}$ & ------ \\
\hline %
\multirow{2}{*}{$N^{*}_{3/2^+}(1900)$} & 0.180 & ------ & 2.422$\times10^{-5}$ \\
\cline{2-4}
 & 0.498 & ------  & 7.744$\times10^{-6}$ \\
\hline %
\multirow{3}{*}{$N^{*}_{1/2^-}(2090)$}& 0.095  & ------
&4.726$\times10^{-5}$\\
\cline{2-4}
 & 0.350  &  ------  & 1.612$\times10^{-5}$ \\
\cline{2-4}
 & 0.414  &  ------  & 2.830$\times10^{-5}$ \\
\hline %
\multirow{3}{*}{$N^{*}_{1/2^+}(2100)$}  & 0.113  & ------ & 1.362$\times10^{-5}$ \\
\cline{2-4}
 & 0.200  & ------   & 2.290$\times10^{-5}$ \\
\cline{2-4}
 & 0.260  & ------   & 2.031$\times10^{-5}$ \\
\hline %
\end{tabular}
 \label{tab:nnpsi}
\end{table}
From this table, we find that $g^{2}_{\psi N N^{*}(1535)}$ is in the
order of $10^{-6}$. Based on the ranges of the measured total width
and the obtained $g^{2}_{\pi N N^{*}}$ for the $N^*_{3/2^+}(1900)$,
$N^*_{1/2^-}(2090)$, and $N^*_{1/2^+}(2100)$ states, the extracted
values of $g^{2}_{\psi N N^{*}(1900)}$, $g^{2}_{\psi N
N^{*}(2090)}$, and $g^{2}_{\psi N N^{*}(2100)}$ could vary in the
ranges of $(0.77\sim2.4)\times10^{-5}$, $(1.6\sim4.7)\times10^{-5}$,
and $(1.4\sim2.3)\times10^{-5}$, respectively. It seems that the
couplings of $J/\psi$ to $N$ and different $N^*$ are about the same.
This is understandable, because that $J/\psi$ is merely composed of
charmed quarks, $N$ is consist of upper and down quarks only, and
$N^*$ is made up of upper, down and even strange quarks, thus the
coupling mechanisms for different $N^*$s would be the same.

\begin{figure}[htbp]
\centering
\subfigure[Type I ($\Gamma_{N^{*}(2090)}$=95MeV).]{
\label{fig:subfig:a} 
\includegraphics[width=6.5cm]{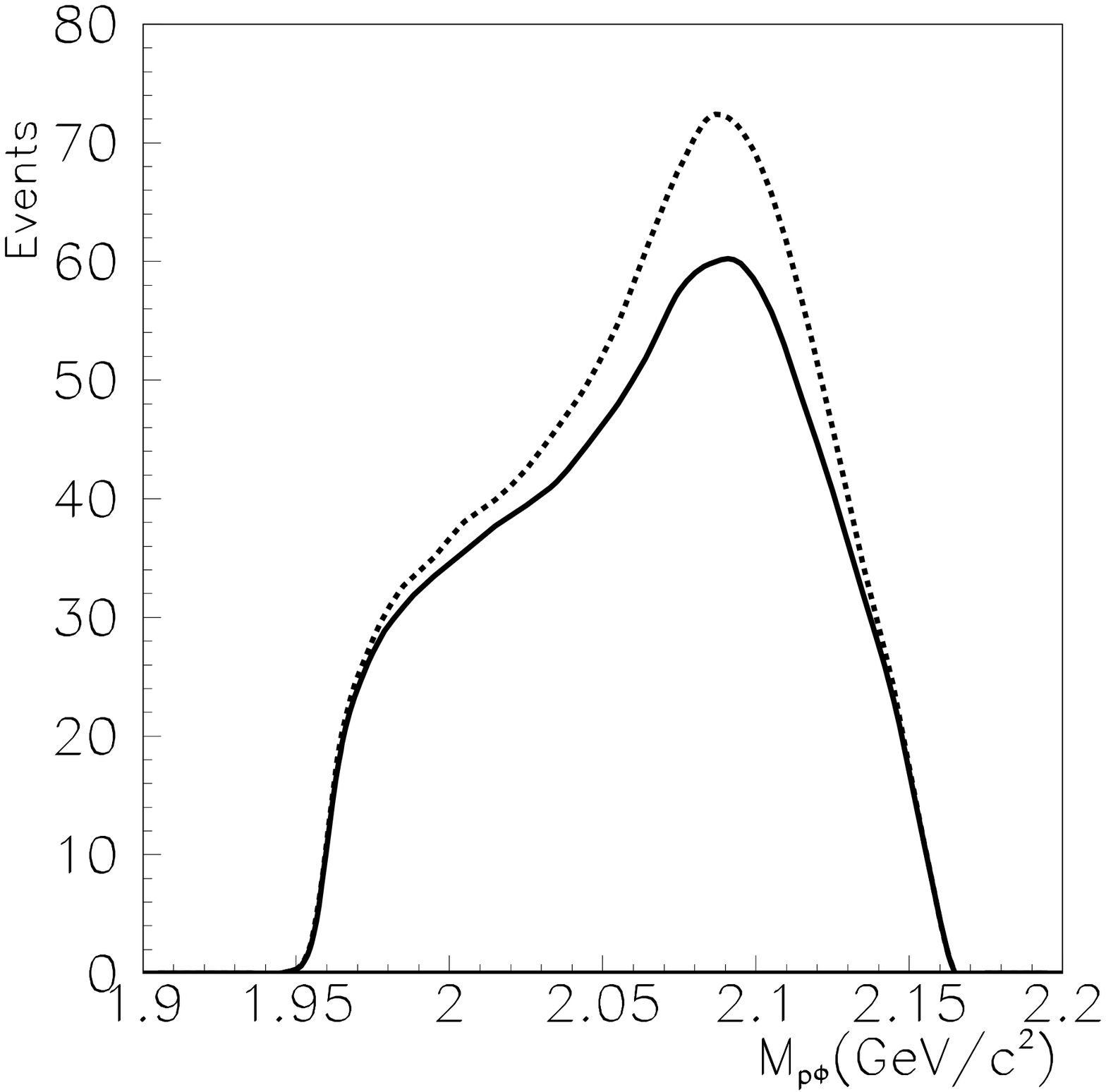}}
\subfigure[Type II ($\Gamma_{N^{*}(2100)}$=113MeV).]{
\label{fig:subfig:b} 
\includegraphics[width=6.5cm]{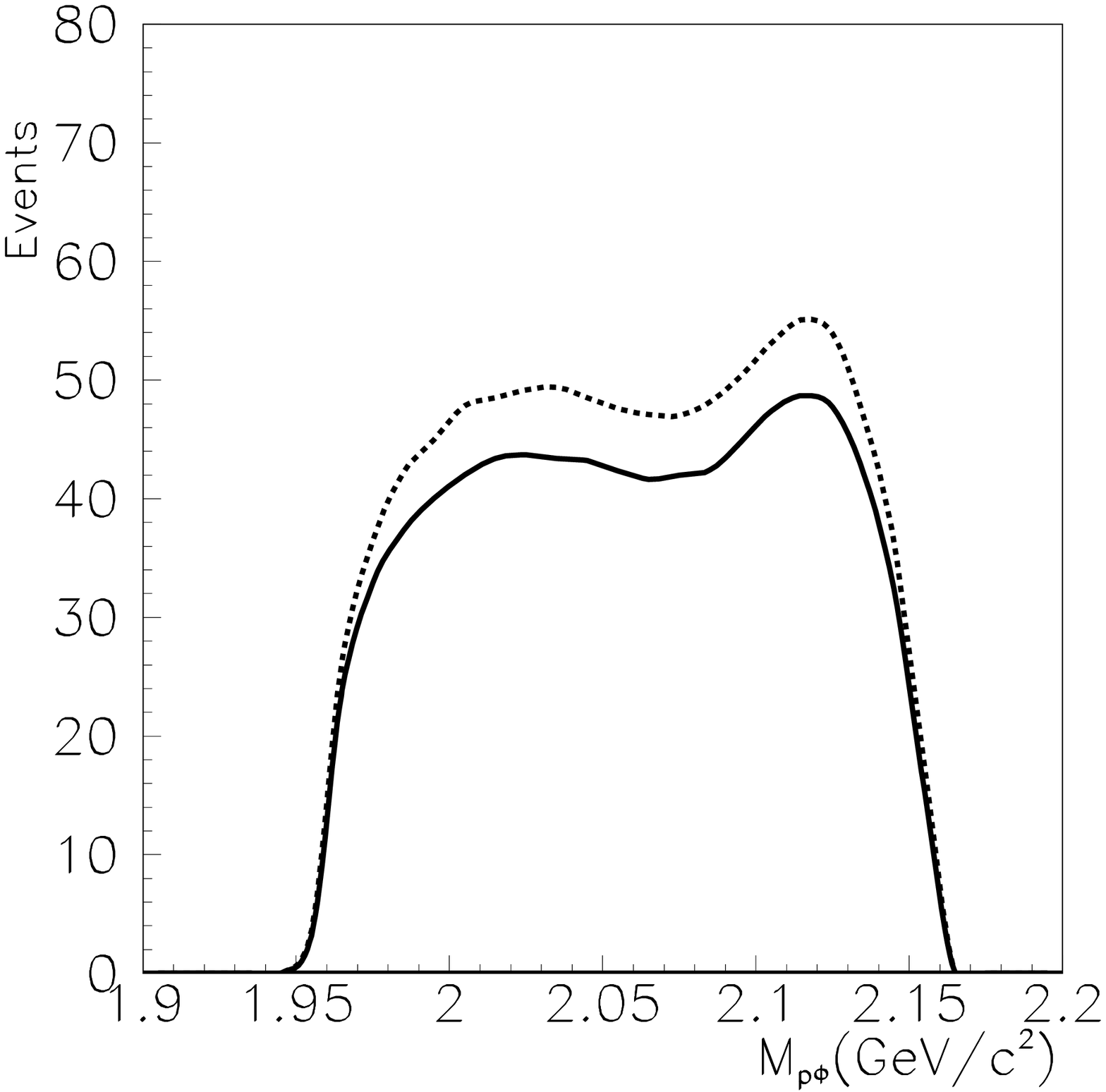}}
\subfigure[Type I
($\Gamma_{N^{*}(1900)}$/$\Gamma_{N^{*}(2090)}$/$\Gamma_{N^{*}(2100)}$
=498MeV/95MeV/260MeV).]{
\label{fig:subfig:a} 
\includegraphics[width=6.5cm]{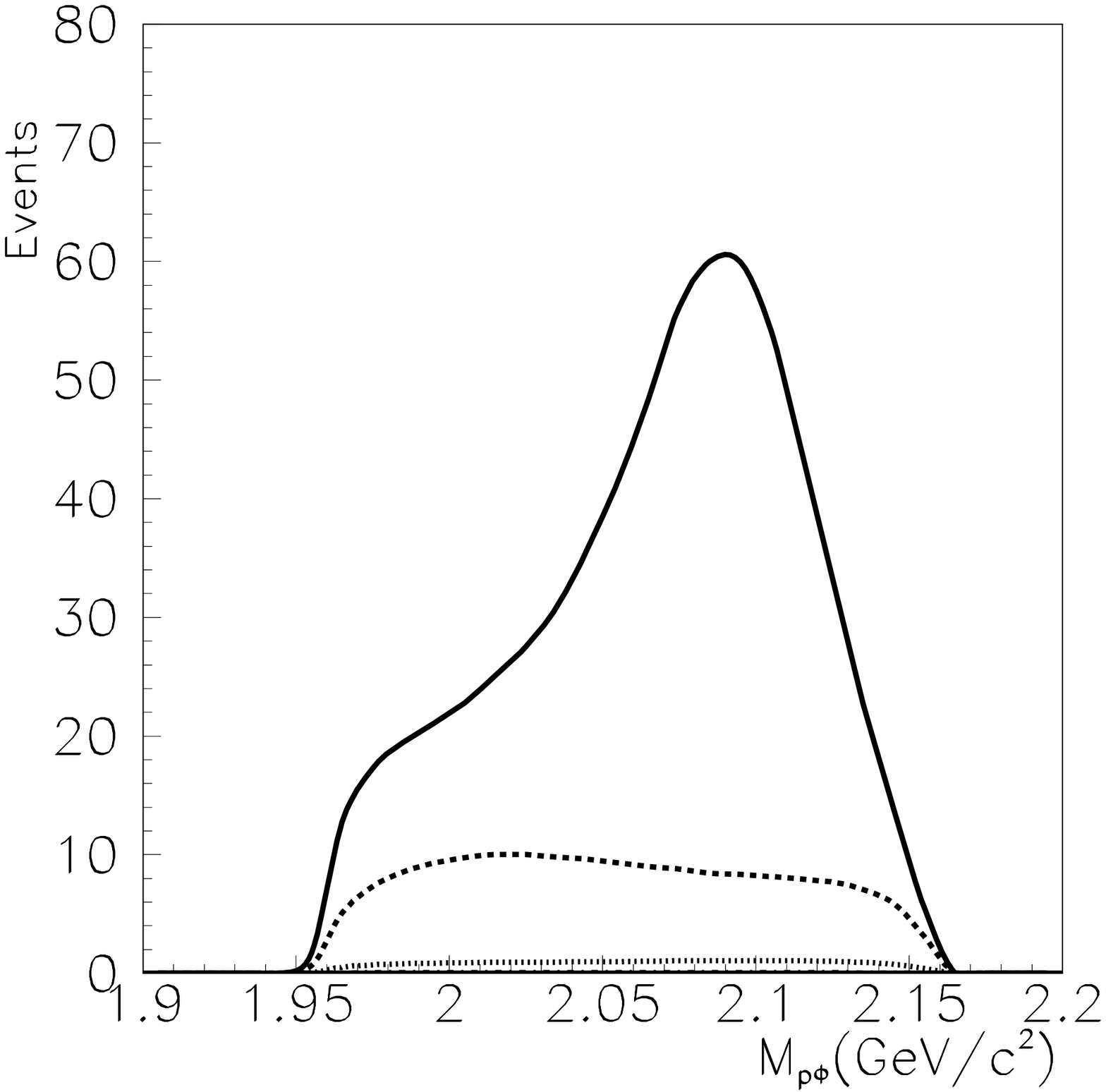}}
\subfigure[Type II
($\Gamma_{N^{*}(1900)}$/$\Gamma_{N^{*}(2090)}$/$\Gamma_{N^{*}(2100)}$
= 180MeV/350MeV/113MeV).]{
\label{fig:subfig:b} 
\includegraphics[width=6.5cm]{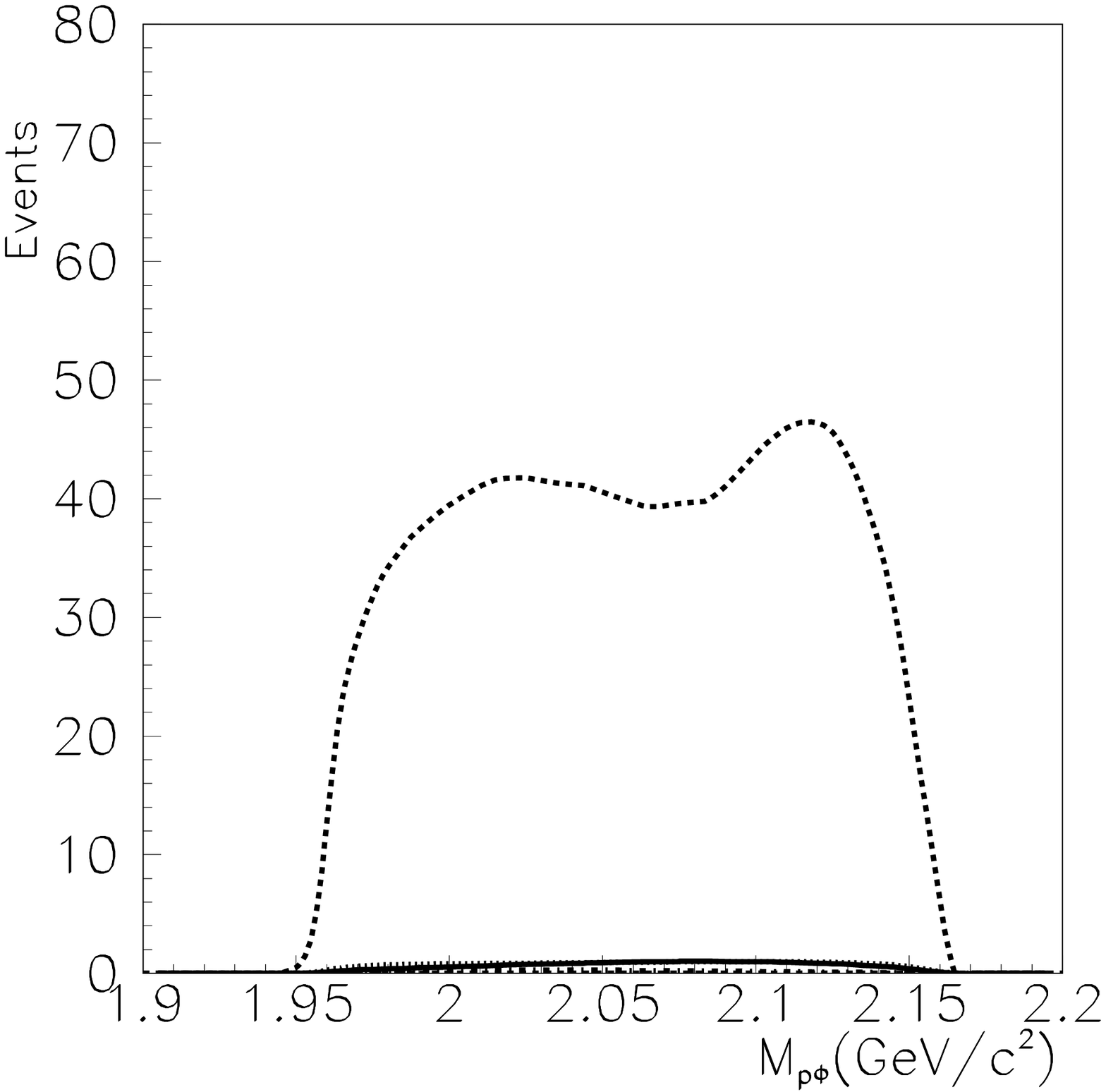}}
\caption{Invariant mass spectra of $p\phi$ in type I
($\Gamma_{N^{*}(2100)}$=113MeV) and type II
($\Gamma_{N^{*}(2090)}$=95MeV) with two curves covering the range of
4 sets of parameters. In (c) and (d), the solid-line is from
$N^{*}(2090)$'s contribution; the dash-line is from $N^{*}(2100)$'s;
the dash-dotted line is from $N^{*}(1535)$'s; the dash-dotted-dotted
line is from $N^{*}(1900)$'s.}
\label{mppphi_SI} 
\end{figure}

In terms of the extracted the values of $g^{2}_{\psi N N^{*}}$ and
$g^{2}_{\phi N N^{*}}$, we are in the stage of calculating physics
observables in the $J/\psi\rightarrow p\bar{p}\phi$ decay with
adopted intermediate states $N^{*}_{1/2^-}(1535)$,
$N^{*}_{3/2^+}(1900)$, $N^{*}_{1/2^-}(2090)$ and
$N^{*}_{1/2^+}(2100)$. The resultant invariant mass spectra of
$p\phi$ are plotted in Fig.~\ref{mppphi_SI}. In sub-figures (a) and
(b), the dashed and solid curves represent the upper and lower
limits of the total invariant mass spectrum, which are caused by the
uncertainties of the widths of the $N^{*}_{3/2^+}(1900)$,
$N^{*}_{1/2^-}(2090)$ and $N^{*}_{1/2^+}(2100)$ states. And in
sub-figures (c) and (d), the dashed, dotted, dash-dotted and
dash-double-dotted curves describe the sub-contributions from the
$N^{*}_{1/2^-}(1535)$, $N^{*}_{3/2^+}(1900)$, $N^{*}_{1/2^-}(2090)$
and $N^{*}_{1/2^+}(2100)$ states, respectively.
The fractions of the contributions for these $N^*$s in the decay are
tabulated in Table~\ref{tab:contribution}.
\begin{table}[htbp]
\centering \caption{Fractions of contributions from
$N^{*}_{1/2^-}(1535)$, $N^{*}_{3/2^+}(1900)$,
$N^{*}_{1/2^-}(2090)$ and $N^{*}_{1/2^+}(2100)$ in the
$J/\psi\rightarrow p\bar{p}\phi$ decay.}
 \label{tab:contribution}
\begin{tabular}{|c|c|c|c|c|c|} \hline
&\multirow{2}{*}{$\Gamma_{N^{*}(1900)}$/ $\Gamma_{N^{*}(2090)}$/
$\Gamma_{N^{*}(2100)}$}  &\multicolumn{4}{|c|}{fraction$(\%)$}
\\ \cline{3-6}
 & &  $N^{*}(1535)$& $N^{*}(1900)$ & $N^{*}(2090)$
&$N^{*}(2100)$ \\
\hline
\multirow{4}{*}{type I}&{180MeV/95MeV/200MeV} & 2.01  &0.44 & 48.22 &32.80 \\
\cline{2-6} &{498MeV/95MeV/200MeV} & 1.96  &0.01 & 61.24 &25.23\\
\cline{2-6}
&{180MeV/95MeV/260MeV} & 1.83  &0.20 & 70.51&15.12 \\
\cline{2-6}
&{498MeV/95MeV/260MeV} & 1.81  &0.01 & 65.36 &16.38 \\
\hline
\multirow{4}{*}{type II}&{180MeV/350MeV/113MeV} & 1.66  &0.36 & 1.42 &76.26 \\
\cline{2-6} &{498MeV/350MeV/113MeV} & 1.65  &0.01 & 1.83 &80.03
\\ \cline{2-6}
&{180MeV/414MeV/113MeV} & 1.69  &0.18 & 2.72&82.38 \\
\cline{2-6}
&{498MeV/414MeV/113MeV} & 1.63  &0.01 & 2.53 &87.56 \\
\hline
\end{tabular}

\end{table}

From the numerical values in Table~\ref{tab:contribution} and the
$p\phi$ invariant mass curves in Fig.~\ref{mppphi_SI}, we have
following observations. From Fig.~\ref{mppphi_SI}(a) and
Table~\ref{tab:contribution}, one sees that in type I the
contribution from $N^{*}_{1/2^-}(2090)$ is about (46.3$\sim$72.1)\%,
and there is a peak structure around 2.09GeV. The sub-contributions
from various $N^*$s shown in Fig.~\ref{mppphi_SI}(c) tell us that
this structure is mainly contributed by $N^{*}_{1/2^-}(2090)$ due to
its relatively narrow width, namely a stronger coupling between $N$
and $\phi$. This implies that there may exists a large $N\phi$ or
$qqqs\bar s$ component in $N^{*}_{1/2^-}(2090)$. Meanwhile the
$N^{*}_{1/2^+}(2100)$ state also provides a sizable contribution of
about (15.1$\sim$32.8)\%, but this contribution is smaller than that
offered by $N^{*}_{1/2^-}(2090)$, and the shape of the contribution
is flatter due to a large width of the state. Therefore, this piece
of contribution would not affect the shape of the total contribution
qualitatively. The contributions from $N^{*}_{1/2^-}(1535)$ and
$N^{*}_{3/2^+}(1900)$ are negligibly small, their contributions are
about (1.81$\sim$2.00)\% and (0.01$\sim$0.44)\%, respectively. The
interference terms can only provide about 2\% of the contribution.
Therefore, disregarding the contributions from the
$N^{*}_{1/2^-}(1535)$ and $N^{*}_{3/2^+}(1900)$ states and the
interference terms will not affect the conclusion qualitatively.
From Fig.~\ref{mppphi_SI}(b) and Table~\ref{tab:contribution}, one
finds that in type II the contribution from $N^{*}_{1/2^+}(2100)$ is
about (76.3$\sim$87.6)\%, and there is also a small peak structure
around 2.11GeV. The sub-contributions plotted in
Fig.~\ref{mppphi_SI}(d) show that this structure almost entirely
comes from contribution of $N^{*}_{1/2^+}(2100)$, because of its
dominant contribution and relatively narrow width. This also implies
that its coupling to $N\phi$ could be remarkable, a significant
$N\phi$ or $qqqs\bar s$ component may exist in
$N^{*}_{1/2^+}(2100)$. Meanwhile the contributions from
$N^{*}_{1/2^-}(1535)$, $N^{*}_{3/2^+}(1900)$ and
$N^{*}_{1/2^-}(2090)$ are negligibly small, their contributions are
about (1.63$\sim$1.70)\%, (0.01$\sim$0.36)\% and (1.33$\sim$2.65)\%,
respectively. The interference terms can only give a contribution
about 2\%. It also shows that one would not be able to explore the
possible strange structures for $N^{*}_{1/2^-}(1535)$ and
$N^{*}_{3/2^+}(1900)$ in the type I case and for
$N^{*}_{1/2^-}(1535)$, $N^{*}_{3/2^+}(1900)$ and
$N^{*}_{1/2^-}(2090)$ in the type II case in this decay process,
because their informations are deeply submerged in the signals of
the $N^{*}_{1/2^-}(2090)$ and $N^{*}_{1/2^+}(2100)$ states and the
$N^{*}_{1/2^+}(2100)$ state, respectively.

\begin{figure}
\centering \subfigure[Type I
($\Gamma_{N^{*}(1900)}$/$\Gamma_{N^{*}(2090)}$/$\Gamma_{N^{*}(2100)}$=
498MeV/95MeV/260MeV.)]{
\label{fig:subfig:b} 
\includegraphics[width=6cm]{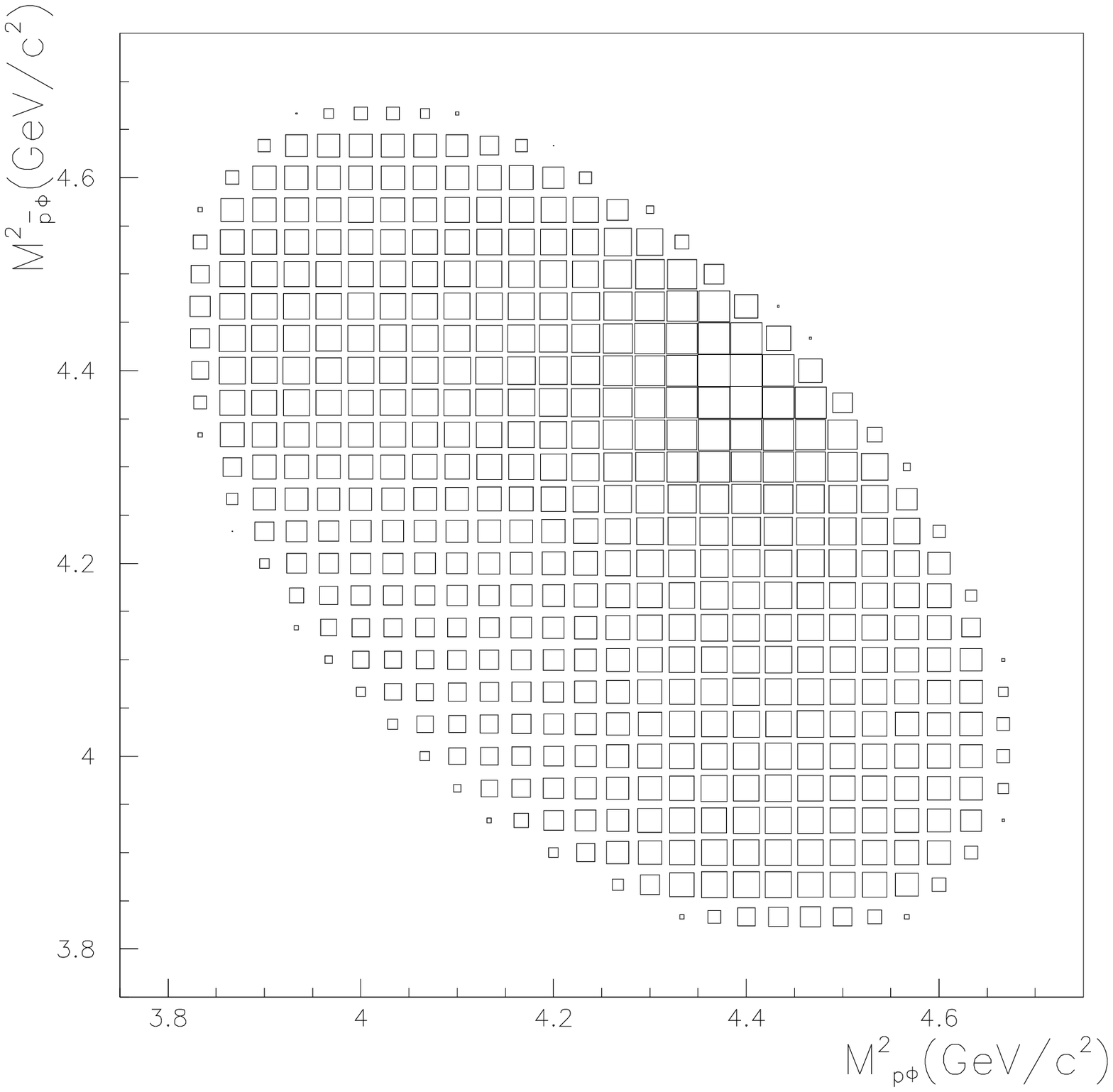}}
\label{fig:subfig:e} \subfigure[Type II
($\Gamma_{N^{*}(1900)}$/$\Gamma_{N^{*}(2090)}$/$\Gamma_{N^{*}(2100)}$=
180MeV/350MeV/95MeV.)]{
\label{fig:subfig:c} 
\includegraphics[width=6cm]{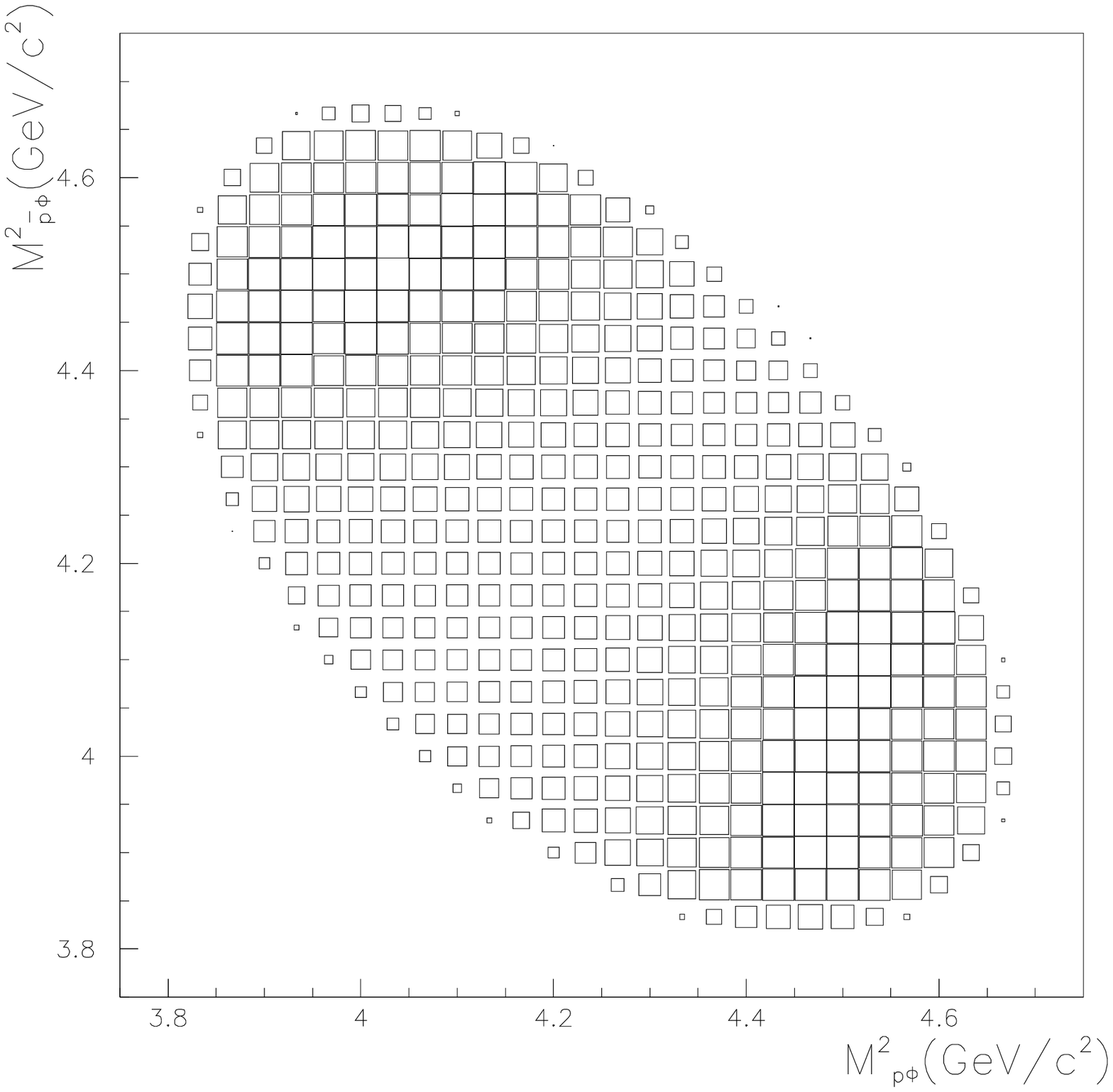}}
\caption{Dalitz plots.}
\label{ppbarphi_dalitz} 
\end{figure}

Furthermore, the Dalitz plots for type I and type II are plotted in
Fig.\ref{ppbarphi_dalitz}. It is shown that the Dalitz plots of
types I and II have distinguishable features. In the type I case,
there are one vertical belt and one horizontal belt at
4.37$(GeV/c^{2})^{2}$ and an enhancement in the upper right corner.
But in the type II case, there is only two enhancements at the upper
left and lower right corners. These patterns agree with the findings
from the invariant mass curves.

Finally, we need to mention that the value of the cut-off parameter
in a certain range does not qualitatively affect our conclusion.

\section{Summary}
In this paper, the $J/\psi\rightarrow p\bar{p}\phi$ decay is studied
in the isobar resonance model with effective Lagrangians. In such a
model, the nucleon resonances are adopted as the intermediate
states. Because of the $s\bar{s}$ structure of the $\phi$-meson and
the OZI rule, the major decay width of this process will be
contributed by the resonances who contain strange content.
Therefore, this decay process could be used to study the possible
strange structure of the nucleon resonances.

Based on a careful analysis, four $N^*$ states,
$N^{*}_{1/2^-}(1535)$, $N^{*}_{3/2^+}(1900)$, $N^{*}_{1/2^-}(2090)$
and $N^{*}_{1/2^+}(2100)$, are adopted in the calculation, so that
the qualitative conclusion would not be affected. The coupling
constants $g^2_{\pi NN^{*} }$ for these $N^*$s and $g^2_{\eta NN^{*}
}$ for $N^{*}_{1/2^-}(1535)$ are extracted from the branching
fractions of the $N^*$s to the $N\pi$ channel and of
$N^{*}_{1/2^-}(1535)$ to the $N\eta$ channel, respectively, in the
first step. With determined $g^2_{\pi NN^{*} }$, coupling constant
$g^2_{\phi NN^{*} }$ for $N^*$s are obtained by fitting the cross
section of the $\pi^{-} p \to n\phi$ reaction. Because the
uncertainties of the partial width for $N^{*}_{3/2^+}(1900)$,
$N^{*}_{1/2^-}(2090)$ and $N^{*}_{1/2^+}(2100)$, the resultant
$g^2_{\phi NN^{*} }$s are allowed to change in certain regions. It
is found that in the best fit, except the dominant contribution from
$N^{*}_{1/2^-}(1535)$ and negligible contribution from
$N^{*}_{3/2^+}(1900)$, the contributions from $N^{*}_{1/2^-}(2090)$
and $N^{*}_{1/2^+}(2100)$ are visible and even remarkable in some
cases, and the total widths of these two states cannot be large
simultaneously. Therefore, there are two types of fits. In the first
type, type I, $N^{*}_{1/2^-}(2090)$ and $N^{*}_{1/2^+}(2100)$ have a
smaller total width and a larger total width, respectively, and in
the second type, type II, it is the other way round. In the second
step, the coupling constant $g^2_{\psi NN^{*}(1535)}$ and $g^2_{\psi
NN^{*}}$ for other three $N^*$s are extracted by fitting the partial
decay widths of the $J/\psi\rightarrow p \bar{p}\eta$ process and
the $J/\psi\rightarrow p \bar{n}\pi^{-}$ process, respectively.

Finally, we can calculate the physical observables in the $J/\psi\to
p \bar{p}\phi$ decay by using obtained $g^2_{\pi NN^{*}}$s and
$g^2_{\psi NN^{*}}$s in the type I and type II cases. The invariant
mass spectrum of $p\phi$ in the type I case shows that there is a
peak structure around 2.09GeV due to the major contribution from the
narrower $N^{*}_{1/2^-}(2090)$ state. This means that its coupling
to $N\phi$ is relatively strong, and a large $N\phi$ or $qqqs\bar s$
component may exist in $N^{*}_{1/2^-}(2090)$. Meanwhile the
contribution from $N^{*}_{1/2^+}(2100)$ is flatter and smaller,
which implies that even there is a strange ingredient in this state,
its coupling to $N\phi$ would be weaker. In the type II case, the
curve of the invariant mass spectrum of $p\phi$ has a small peak
structure around 2.11GeV, because of the dominant contribution from
the narrow $N^{*}_{1/2^+}(2100)$ state and negligible contributions
from other states. It suggests that its coupling to $N\phi$ is
strong, a significant $N\phi$ or $qqqs\bar s$ component might exist
in the $N^{*}_{1/2^+}(2100)$. However, one would not be able to
reveal the strange structure in $N^{*}_{1/2^-}(2090)$, because its
information is deeply submerged in the signal of the
$N^{*}_{1/2^+}(2100)$ state. For the same reason, no matter in which
cases, one cannot figure out strange structures of
$N^{*}_{1/2^-}(1535)$ and $N^{*}_{3/2^+}(1900)$ from this process.

In summary, in the $J/\psi\to p \bar{p}\phi$ decay, the widths of
$N^{*}_{1/2^-}(2090)$ and $N^{*}_{1/2^+}(2100)$ cannot be large
simultaneously. The proposed study of this channel with the high
statistics BESIII data~\cite{BES-yb} will tell us how the $p\phi$
invariant mass curve goes. If the shape of the curve likes that of
type I, the width of the $N^{*}_{1/2^-}(2090)$ state is narrower and
there would be a considerable mount of $p\phi$ or $qqqs\bar s$
component in the state, while the width of the $N^{*}_{1/2^+}(2100)$
state would be wider. If the shape of the curve is similar to that
of type II, only the width of the $N^{*}_{1/2^+}(2100)$ state is
narrower and there would be a certain mount of $p\phi$ or $qqqs\bar
s$ component in the state. Of course, the real data of high
statistics on the $J/\psi\to p \bar{p}\phi$ decay may reveal more
knowledge on all possible $N^*$s than our predictions based on the
information from $\pi N\to\phi N$. It will definitely provide us
useful information on the $N^*$ resonances with large $qqqs\bar s$
component. And the $pp\to
pp\phi$ reaction should also be studied to confirm our prediction.\\

\textbf{Acknowledgments}\\
This work is partly supported by the National Natural Science
Foundation of China under grants Nos. 10875133, 10847159, 10975038,
11035006, 11165005, and the Key-project by the Chinese Academy of
Sciences under project No. KJCX2-EW-N01, and the Ministry of Science
and Technology of China (2009CB825200).


\newpage
\begin{appendix}
\begin{center}
\textbf{APPENDIX}
\end{center}
\section{}
The invariant amplitudes of $\pi^{-} p\rightarrow n\phi$ reaction
with $N^{*}(1535)$, $N^{*}(1900)$, $N^{*}(2090)$ and $N^{*}(2100)$
being intermediate states are as follows: \\
For $N^{*}(1535)$
\begin{eqnarray}
{\cal M}_{N^*(1535)}=\sqrt{2} g_{\pi N N^{*}}g_{\phi N
N^{*}}F_{N^{*}}(q^{2})\bar{u}(p_{n},s_{n})\gamma_{5}(\gamma_{\nu}-
\frac{q_{\nu}\not\!
q}{q^{2}})\epsilon^{\nu}(p_{\phi},s_{\phi})G_{N^{*}}(q)u(p_{p},s_{p}),
\end{eqnarray}
for $N^{*}(1900)$
\begin{eqnarray}
{\cal M}_{N^*(1900)}=\frac{i\sqrt{2}}{M_{N^{*}}}g_{\pi N
N^{*}}g_{\phi N N^{*}}F_{N^{*}}(q^{2})\bar{u}(p_{n},s_{n})\gamma_{5}
\epsilon_{\mu}(p_{\phi},s_{\phi})G^{\mu\nu}_{N^{*}}(q)
p_{\pi\nu}u(p_{p},s_{p}),
\end{eqnarray}
For $N^{*}(2090)$
\begin{eqnarray}
{\cal M}_{N^*(2090)}=\sqrt{2}g_{\pi N N^{*}}g_{\phi N
N^{*}}F_{N^{*}}(q^{2})\bar{u}(p_{n},s_{n})\gamma_{5}(\gamma_{\nu}-
\frac{q_{\nu}\not\!
q}{q^{2}})\epsilon^{\nu}(p_{\phi},s_{\phi})G_{N^{*}}(q)u(p_{p},s_{p}),
\end{eqnarray}
and for $N^{*}(2100)$
\begin{eqnarray}
{\cal M}_{N^*(2100)}=i\sqrt{2} g_{\pi N N^{*}}g_{\phi N
N^{*}}F_{N^{*}}(q^{2})\bar{u}(p_{n},s_{n})\gamma_{\nu}\epsilon^{\nu}(p_{\phi},s_{\phi})G_{N^{*}}(q)\gamma_{5}u(p_{p},s_{p}),
\end{eqnarray}
where $p_{\pi}$ is the four momentum of $\pi^{-}$ meson, and
$\epsilon(p_{\phi})$ is the polarization vector of $\phi$ meson.

\section{}
The invariant amplitude of $J/\psi\rightarrow p\bar{p}\eta$ decay
with $N^{*}(1535)$ being the intermediate state is written as:
\begin{eqnarray}\nonumber
{\cal M}_{N^{*}(1535)}=&&i g_{\eta N N^{*}}g_{\psi N
N^{*}}\bar{u}(p_{1},s_{1})[G_{N^{*}}(q)F_{N}(q^{2}
)\gamma_{5}\sigma_{\mu\rho}p^{\rho}_{\psi}\epsilon^{\mu}(k,s_{\psi}) \\
 && +
\gamma_{5}\sigma_{\mu\rho}p^{\rho}_{\psi}\epsilon^{\mu}(k,s_{\psi})
G_{\bar{N}^{*}}(q')F_{N}(q'^{2})]v(p_{2},s_{2}),
\end{eqnarray}
with $p_{\psi}$ and $\epsilon(p_{\psi})$ being the four momentum and
the polarization vector of $J/\psi$.

The invariant amplitudes of $J/\psi\rightarrow p\bar{n}\pi^{-}$
reaction with $N^{*}(1900)$, $N^{*}(2090)$ and $N^{*}(2100)$ being
intermediate states are as follows:\\
For $N^{*}(1900)$,
\begin{eqnarray}
{\cal M}_{N^*(1900)}=\frac{i \sqrt{2}}{M_{N^{*}}}g_{\pi N
N^{*}}g_{\psi N
N^{*}}\bar{u}(p_{1},s_{1})\gamma_{5}\epsilon_{\mu}(k,s_{\psi})
G^{\mu\nu}_{\bar{N}^{*}}(q')F_{N}(q'^{2})p_{\pi\nu}v(p_{2},s_{2}).
\end{eqnarray}
For $N^{*}(2090)$,
\begin{eqnarray}
{\cal M}_{N^*(2090)}=i \sqrt{2}g_{\pi N N^{*}}g_{\psi N
N^{*}}\bar{u}(p_{1},s_{1})
\gamma_{5}\sigma_{\mu\rho}p^{\rho}_{\psi}\epsilon^{\mu}(k,s_{\psi})
G_{\bar{N}^{*}}(q')F_{N}(q'^{2})v(p_{2},s_{2}).
\end{eqnarray}
For $N^{*}(2100)$,
\begin{eqnarray}
{\cal M}_{N^*(2100)}=i \sqrt{2}g_{\pi N N^{*}}g_{\psi N
N^{*}}\bar{u}(p_{1},s_{1})
\gamma_{\mu}\epsilon^{\mu}(k,s_{\psi})G_{\bar{N}^{*}}(q')F_{N}(q'^{2})\gamma_{5}v(p_{2},s_{2}).
\end{eqnarray}

\section{}
The invariant amplitudes of $J/\psi\rightarrow p\bar{p}\phi$ decay
with $N^{*}(1535)$, $N^{*}(1900)$, $N^{*}(2090)$ and
$N^{*}(2100)$ being intermediate states are as follows:\\
 For $N^{*}(1535)$,
\begin{eqnarray}\nonumber
{\cal M}_{N^*(1535)}=&&i g_{\phi N N^{*}}g_{\psi N
N^{*}}\bar{u}(p_{1},s_{1})[\gamma_{5}(\gamma_{\nu}-
\frac{q_{\nu}\not\!q
}{q^{2}})\epsilon^{\nu}(p_{3},s_{\phi})G_{N^{*}}(q)F_{N}(q^{2}
)\gamma_{5}\sigma_{\mu\rho}p^{\rho}_{\psi}\epsilon^{\mu}(k,s_{\psi}) \\
 && +
\gamma_{5}\sigma_{\mu\rho}p^{\rho}_{\psi}\epsilon^{\mu}(k,s_{\psi})
G_{\bar{N}^{*}}(q')F_{N}(q'^{2})\gamma_{5}(\gamma_{\nu}-
\frac{q'_{\nu}\not\!
q'}{q'^{2}})\epsilon^{\nu}(p_{3},s_{\phi})]v(p_{2},s_{2}).
\end{eqnarray}
For $N^{*}(1900)$,
\begin{eqnarray}\nonumber
{\cal M}_{N^*(1900)}=&&-g_{\phi N N^{*}}g_{\psi N
N^{*}}\bar{u}(p_{1},s_{1})[\gamma_{5}\epsilon_{\nu}(p_{3},s_{\phi})G^{\nu\mu}_{N^{*}}(q)F_{N}(q^{2}
)\gamma_{5}\epsilon_{\mu}(k,s_{\psi}) \\
 && + \gamma_{5}\epsilon_{\mu}(k,s_{\psi})
G^{\mu\nu}_{\bar{N}^{*}}(q')F_{N}(q'^{2})\gamma_{5}\epsilon_{\nu}(p_{3},s_{\phi})]v(p_{2},s_{2}).
\end{eqnarray}
For $N^{*}(2090)$,
\begin{eqnarray}\nonumber
{\cal M}_{N^*(2090)}=&&i g_{\phi N N^{*}}g_{\psi N
N^{*}}\bar{u}(p_{1},s_{1})[\gamma_{5}(\gamma_{\nu}-
\frac{q_{\nu}\not\!q
}{q^{2}})\epsilon^{\nu}(p_{3},s_{\phi})G_{N^{*}}(q)F_{N}(q^{2}
)\gamma_{5}\sigma_{\mu\rho}p^{\rho}_{\psi}\epsilon^{\mu}(k,s_{\psi}) \\
 && +
\gamma_{5}\sigma_{\mu\rho}p^{\rho}_{\psi}\epsilon^{\mu}(k,s_{\psi})
G_{\bar{N}^{*}}(q')F_{N}(q'^{2})\gamma_{5}(\gamma_{\nu}-
\frac{q'_{\nu}\not\!
q'}{q'^{2}})\epsilon^{\nu}(p_{3},s_{\phi})]v(p_{2},s_{2}).
\end{eqnarray}
For $N^{*}(2100)$,
\begin{eqnarray}\nonumber
{\cal M}_{N^*(2100)}=&&g_{\phi N N^{*}}g_{\psi N
N^{*}}\bar{u}(p_{1},s_{1})[\gamma_{\nu}\epsilon^{\nu}(p_{3},s_{\phi})G_{N^{*}}(q)F_{N}(q^{2}
)\gamma_{\mu}\epsilon^{\mu}(k,s_{\psi}) \\
 && +
\gamma_{\mu}\epsilon^{\mu}(k,s_{\psi})G_{\bar{N}^{*}}(q')F_{N}(q'^{2})\gamma_{\nu}\epsilon^{\nu}(p_{3},s_{\phi})]v(p_{2},s_{2}).
\end{eqnarray}
\end{appendix}



\end{document}